\documentclass[twocolumn,aps,prc]{revtex4}

\usepackage{float}
\usepackage{url}
\usepackage{bm}

\usepackage[utf8]{inputenc}
\usepackage{amsfonts}
\usepackage{amsmath}
\usepackage{amssymb}
\usepackage{epsfig}
\usepackage{bm}

\usepackage{fixme}
\usepackage{color}
\usepackage{cancel}

\graphicspath{{./figures/}}

\begin{document}
\title{Three-body calculations of beta decay applied to $^{11}$Li}

\author{E. Garrido$^1$, A.S. Jensen$^2$, H.O.U. Fynbo$^2$, K. Riisager$^2$}
\affiliation{$^1$Instituto de Estructura de la Materia, CSIC, Serrano 123, E-28006 Madrid, Spain}

\affiliation{$^{2}$Department of Physics and Astronomy, Aarhus University, DK-8000 Aarhus C, Denmark}

\date{\today}

\begin{abstract}
A novel practical few-body method is formulated to include isospin
symmetry for nuclear halo structures.  The method is designed to
describe beta decay, where the basic concept of isospin symmetry
facilitates a proper understanding.  Both isobaric analogue and
anti-analogue states are treated.  We derive general and explicit
formulas for three-body systems using hyperspherical coordinates.  The
example of the beta decaying $^{11}$Li ($^{9}$Li+$n$+$n$) is chosen as a
challenging application for numerical calculations of practical
interest.  The detailed results are compared to existing experimental
data and good agreement is found at high excitation energies, where
the isobaric analogue and anti-analogue states are situated in the
daughter nucleus.  An interpretation of the decay pattern at lower
excitation energies is suggested.  Decays of the $^{9}$Li-core and the
two halo-neutrons are individually treated and combined to the
daughter system with almost unique isospin, which we predict to be
broken by about $0.4\%$ probability.  Properties of decay products are
predicted as possible future tests of this model.
\end{abstract}

\maketitle

\section{Introduction}
Nuclear halo states are characterized by a significant decoupling of the halo particle(s) from the remainder of the nucleus \cite{Jen05,Rii13}. Many accounts of halo nuclei focus on the spatial decoupling of the halo and the core, but the decoupling of dynamic degrees of freedom is of similar importance. This has implications for many aspects of their structure and for the dynamics when undergoing nuclear reactions. The aim of the current paper is to explore how beta decay of halo nuclei is affected through the use of a few-body model that enables treatment also of the continuum behaviour.

Most established halo states are ground states of light nuclei in the
vicinity of the neutron or proton driplines. Here beta decay energies
are large, and allowed transitions dominate the decay pattern, see
\cite{Bla08,Pfu12} for general overviews. Beta decays are a valuable
probe of nuclear structure with operators that for allowed transitions do not involve the spatial coordinates. The decays of halo nuclei are reviewed in a recent paper \cite{Rii22}, and two specific features are important for the current work. The first is that the large spatial extension of the halo wave functions can give rise to decays that are most efficiently described as proceeding directly to continuum states, as done naturally in few-body models. The second is that the decoupling between halo and core may lead to distinct patterns in the decay as discussed in the next section.
It should be recalled that the physics structure is often seen best in transitions of large beta strength that only rarely correspond to those with high branching ratio.

The nucleus $^{11}$Li is not only one of the best studied halo nuclei, but also has a very diverse break-up pattern during the beta decay process \cite{Rii22}, and it is therefore natural to take a close look at this decay. Furthermore, it has been
difficult to pin down the many decay branches of the nucleus, not only experimentally, but also from the theoretical point of view. This is connected to the fact that most of the beta strength in the $^{11}$Li decay leads to break-up into continuum states.
For example, with the shell model \cite{Suz97,Mad09a} or the antisymmetrized molecular dynamics \cite{Kan10}, the calculations are often restricted to the beta strength distribution in $^{11}$Be, and they do not attempt to describe the subsequent particle emission or the coupling to the continuum. The only exception is calculations \cite{Ohb95,Zhu95,Bay06} of the beta-delayed deuteron branch that assume decays direct to the continuum.

In this work, we shall investigate the beta decay of the $^{11}$Li
nucleus in its ground state.  The calculations will be based on a
three-body framework (core plus two halo nucleons), by means of the
hyperspherical adiabatic expansion method \cite{nie01}.  Earlier the
method was used to describe not only $^{11}$Li, but also its mirror
nucleus $^{11}$O, the two extremes of a $T=5/2$ isospin multiplet
\cite{Gar20}.  We now extended this two- and three-body general
formulation to treat the isospin symmetry fully and consistently for
other less extreme members of the multiplet.  The extension to
multiplet members in between, such as the Isobaric Analog State (IAS)
or the Anti-Analog State (AAS) in $^{11}$Be will be considered in
detail.  We shall provide the necessary technical developments to be
implemented into the standard three-body method.  The extended method
also allows to determine which asymptotic channels the decays proceed
through.

After this introduction, we start in Section~II by outlining isospin
structures and concepts in the beta decay process that are the basis
of the subsequent sections.  The general numerical three-body method
used in the calculations, and in particular, the new implementation of
the isospin degrees of freedom will be described in Section III.  In
Section~IV, we focus on the particular case of $^{11}$Li,
describing the isospin structure of the wave functions, matrix
elements, and energies.  In Sections~V and VI we continue with the
corresponding numerical calculations of respectively, three-body wave
functions and beta decay strength.  A comparison to the experimentally established 
decay scheme follows in Section~VII. Finally,
we conclude in Section~VIII with a brief outlook. Some technical details
and supplementary information are given in appendices.

\section{Conceptual background}
We use the convention where the neutron has third isospin component $-\frac{1}{2}$. The operators for allowed Fermi and Gamow-Teller $\beta^{\mp}$ decay then have the form $\hat{\cal O}_F^\mp= \sum_i t_i^{\pm} = T^{\pm}$ and $\hat{\cal O}_{GT}^{\mp} = \sum_i \bm{\sigma}_i t_i^{\pm}$; here $\bm{t}_i$ and $\bm{\sigma}_i\hbar/2$ are, respectively, the isospin
and spin operators acting on nucleon $i$, and $\bm{T}$ 
is the total isospin operator of the nucleus. The ladder operator $t_i^+$ is defined as $t_i^+=t_{x_i}+ i t_{y_i}$, where
$t_{x_i}$ and $t_{y_i}$ are the Cartesian $x$- and $y$-components of the vector operator $\bm{t}_i$.

\subsection{Core and halo decays}

Let us consider first the case where a halo state has one or more neutrons in the halo component (the modifications to proton halos are then trivial) and that the mass numbers of the core and halo components are $A_c$ and $A_h$ with total mass  number $A=A_c+A_h$. The total isospin is split into core and halo parts, $\bm{T} = \bm{t_c} + \bm{t_h}$, and we can assume that the quantum numbers fulfill $t_h = A_h/2$, $T=t_c+t_h$, whose respective third components are given by $T_z=-T$, $t_c^z=-t_c$, $t_h^z = -t_h$.

Our basic approximation is taken from \cite{Nil00, Jon04}. It is applicable for pronounced halo states where the wave function can be factorized into a core and a halo part. An allowed beta-decay operator $\hat{\mathcal{O}}_{\beta}^A$, either Fermi
or Gamow-Teller, is a sum over operators for each nucleon. Therefore, one formally can divide the sum into two components, corresponding to core decay and halo decay:
\begin{equation} \label{eq:decay}
   \hat{\cal O}_{\beta}^A |\mathrm{c+h}\rangle =
   \hat{\cal O}_{\beta}^{A}
    \left( |\mathrm{c}\rangle |\mathrm{h}\rangle \right) =
     |\mathrm{c}\rangle
    \left( \hat{\cal O}_{\beta}^{A_h} |\mathrm{h}\rangle \right) +
    \left( \hat{\cal O}_{\beta}^{A_c} |\mathrm{c}\rangle \right)
      |\mathrm{h}\rangle,
\end{equation}
where $|\mathrm{c}\rangle$ and $|\mathrm{h}\rangle$ represent the (decoupled) core and halo parts of the wave function.

As illustrated shortly, the two final components will in general not have an individual well-defined isospin. Furthermore, the core-decay component often contains several terms corresponding to different levels in its daughter nucleus. Therefore, the right-hand side in general does not give eigenstates in the final system, but may suggest a skeleton to be used in interpreting patterns in the beta decay.
Some experimental support for our approximation is found in the decays of the halo nuclei $^{6}$He ($\alpha+2n$) and $^{14}$Be ($^{10}$Be$+4n$) \cite{Rii22,Nil00, Jon04}: in the former case $t_c=0$ and the decay is a pure halo decay, in the latter case the main decay branch (to a low-lying $1^+$ level in $^{14}$B) is very similar to the decay of the $^{12}$Be core so this branch looks like a ``core decay''.

Our procedure will then be to calculate the initial wave function in a
few-body model that includes all halo degrees of freedom, and act on
it with the beta-decay operators according to Eq.\
(\ref{eq:decay}). The resulting wave function then must be projected
on the final states. Our computational challenge is not to find an
accurate halo wave function, but to obtain a realistic description of
the final states. The main simplifying assumption is to assume
few-body structures in the final states also, and calculate these in a
similar manner, but of course with adjusted interactions. This cannot
be expected to reproduce all details of the final states, but may
provide the main features of how the beta strength is distributed in
energy and thus could be a good approximation for the components with large beta strength.

\subsection{Isospin, IAS and AAS}
\label{sec21}

Isospin is a good quantum number for most nuclear states and is particularly relevant for beta decay, since the operator for allowed Fermi decay is simply the isospin raising/lowering operator. We refer to Appendix~\ref{app3} for a more extensive discussion and give here the relevant wave functions for the general (neutron-rich) halo system considered above. The notation reflects that we shall mainly be interested in two-neutron halo nuclei where the isospins of the halo nucleons, $t_2$ and $t_3$, couple to the halo isospin, $t_h$, which in turn couples to the core isospin, $t_c$, to give the total isospin, $T$, with projection $T_z$. 

The isospin wave functions are then for the halo state, the isobaric analogue state (IAS),
and the corresponding anti-analogue state (AAS), given by:
\begin{eqnarray}
|\mathrm{halo}\rangle &=&
|(t_2, t_3)t_h, t_c ; T=t_h+t_c,T_z=-t_h-t_c\rangle  \nonumber \\ &=&
|t_c,t_c^z=-t_c\rangle |t_h,t_h^z=-t_h\rangle,
\label{isohalogen}
\end{eqnarray}
\begin{eqnarray}
| \mathrm{IAS} \rangle&=&|(t_2,t_3)t_h,t_c ; T=t_h+t_c,T_z=-t_h-t_c+1\rangle \nonumber \\  &=&
\sqrt{\frac{t_h}{t_c+t_h}}  |t_c,t_c^z=-t_c\rangle |t_h,t_h^z=-t_h+1\rangle   \nonumber \\ &+&
 \sqrt{\frac{t_c}{t_c+t_h}}  |t_c,t_c^z=-t_c+1\rangle |t_h,t_h^z=-t_h\rangle,
\label{isoIASgen} 
\end{eqnarray}
and 
\begin{eqnarray}
| \mathrm{AAS} \rangle&=& 
  |(t_2,t_3)t_h,t_c; T=t_h+t_c-1,T_z=-t_h-t_c+1\rangle \nonumber \\  &=&
 \sqrt{\frac{t_c}{t_c+t_h}}  |t_c,t_c^z=-t_c\rangle |t_h,t_h^z=-t_h+1\rangle  \nonumber \\ &-& 
 \sqrt{\frac{t_h}{t_c+t_h}}  |t_c,t_c^z=-t_c+1\rangle |t_h,t_h^z=-t_h\rangle,
\label{isoAASgen}
\end{eqnarray}
where $t_c^z$ and $t_h^z$ are the projections of $t_c$ and $t_h$,
respectively. If $t_c=0$ only the IAS exists and it is given by the
first term in Eq. (\ref{isoIASgen}). Note that whereas the IAS
normally remains quite pure, the AAS will often be close to states in the
daughter nucleus with the same (lower) isospin, spin and parity and may mix
with those states.

\section{Numerical three-body method}

In this work we assume that the nuclei involved in our investigation can be described as a core surrounded by two halo nucleons. Therefore, together with the three-body
method itself, all the required internal two-body interactions need to be specified. On top of that, the decays of 
the core and nucleon halo have to be both considered.

\subsection{Formulation}

The total three-body wave function describing the system will be obtained by means of the hyperspherical adiabatic expansion method detailed in \cite{nie01}. According to this method the three-body wave functions are written as:
\begin{equation}
 \Psi(\bm{x},\bm{y})=\frac{1}{\rho^{5/2}}\sum_n f_n(\rho) \Phi_n(\rho,\Omega),
 \label{wf}
\end{equation}
where $\bm{x}$ and $\bm{y}$ are the usual Jacobi coordinates, from which one can define the hyperradius $\rho$ and the five hyperangles (collected into $\Omega$) \cite{nie01}.
The angular functions $\Phi_n(\rho,\Omega)$ are the eigenfunctions of the angular part of the Schr\"{o}dinger, or Faddeev, equations, with eigenvalues $\lambda_n(\rho)$. The radial functions, $f_n(\rho)$,
are obtained as the solution of a coupled set of differential equations in which the angular eigenvalues, $\lambda_n(\rho)$, enter as effective potentials:
\begin{equation}
V_\mathrm{eff}(\rho)=\frac{\hbar^2}{2 m} \frac{\lambda_n(\rho)+\frac{15}{4}}{\rho^2},
\label{veff}
\end{equation}
where $m$ is the arbitrary normalization mass used to construct the Jacobi coordinates \cite{nie01}.

In the calculations, the angular eigenfunctions $\Phi_n$ are expanded in terms of the basis set $\{| {\cal Y}_q\rangle\}$, where ${\cal Y}_q$ represents the coupling  between the usual
hyherspherical harmonics and the spin terms, and where $q$ collects all the necessary, orbital and spin, quantum numbers. 
The number of components, as well as the number of terms in the expansion in Eq.(\ref{wf}), should of
course be large enough to get a converged three-body wave function.

The calculations, as described in Ref.~\cite{nie01}, have now to be extended to make explicit the isospin quantum numbers associated to the system under investigation. As mentioned above, the isospin state of the halo system, the IAS, and the AAS, 
are given by the isospin wave functions specified in
Eqs.(\ref{isohalogen}), (\ref{isoIASgen}), and (\ref{isoAASgen}), respectively. The isospin quantum numbers characterize
each of the states to be computed. With this in mind, the simplest procedure to incorporate the isospin state into the
calculation is to proceed as described in \cite{nie01}, but where the basis set used for the expansion of the angular eigenfunctions, $\Phi_n$, is not just given by the 
$\{|{\cal Y}_q\rangle\}$ terms formed after the coupling of the hyperspherical harmonics and the spin functions, but by  the set  $\{|{\cal Y}_q\rangle |T T_z\rangle\}$,
where $|TT_z\rangle \equiv |(t_2t_3)t_h, t_c; T T_z\rangle$ is the isospin function of the system to be computed. 

In other words, following the procedure sketched above, the full three-body wave function, $\Psi$, is in practice
written as:
\begin{equation}
\Psi=\Psi_{3b} |T T_z\rangle,
\label{wffull}
\end{equation}
where $\Psi_{3b}$ describes the relative motion between the three constituents of the system.

\subsection{Core energy}
\label{coren}

An important point to take into account is that the total hamiltonian is actually the sum of the one describing the relative motion of the three constituents of the system, plus the core hamiltonian,
${\cal H}_{\mathrm{core}}$, which describes the core degrees of freedom. The hamiltonian ${\cal H}_{\mathrm{core}}$ is such that:
\begin{equation}
{\cal H}_\mathrm{core} |t_c,t_c^z \rangle =\xi_\mathrm{core} |t_c,t_c^z \rangle
\label{hcgen}
\end{equation}
where $\xi_\mathrm{core}$ is the energy of the core.

Obviously, in order to obtain the three-body wave function, the diagonalization of the full hamiltonian
will require inclusion of the all the matrix elements of  ${\cal H}_\mathrm{core}$
between all the elements of the basis set used for the expansion, $\{|{\cal Y}_q\rangle |T T_z\rangle\}$.
Making use of Eqs.(\ref{isohalogen}), (\ref{isoIASgen}), and (\ref{isoAASgen}), we can easily see that these 
matrix elements for the halo states, the IAS, and the AAS, are given by:
\begin{equation}
\langle {\cal Y}_q; \mathrm{halo} | {\cal H}_\mathrm{core}  |{\cal Y}_{q'}; \mathrm{halo} \rangle=
 \xi_{-t_c} \delta_{qq'},
 \label{coreh}
\end{equation}
\begin{eqnarray}
\lefteqn{ \hspace*{-1cm}
\langle {\cal Y}_q; \mathrm{IAS} | {\cal H}_\mathrm{core}  |{\cal Y}_{q'}; \mathrm{IAS} \rangle=} \nonumber \\ & &
\left(
\frac{t_h}{t_c+t_h} \xi_{-t_c} +\frac{t_c}{t_c+t_h} \xi_{-t_c+1}
\right) \delta_{qq'},
\label{coreIAS}
\end{eqnarray}
and
\begin{eqnarray}
\lefteqn{\hspace*{-0.5cm}
\langle {\cal Y}_q; \mathrm{AAS} | {\cal H}_\mathrm{core}  |{\cal Y}_{q'}; \mathrm{AAS} \rangle=} \nonumber \\ &&
\left(
\frac{t_c}{t_c+t_h} \xi_{-t_c} +\frac{t_h}{t_c+t_h} \xi_{-t_c+1}
\right) \delta_{qq'},
\label{coreAAS}
\end{eqnarray}
where $\xi_{-t_c}$ and $\xi_{-t_c+1}$ are, respectively, the energies
of the core represented by the $|t_c,t_c^z=-t_c\rangle$ and $|t_c,t_c^z=-t_c+1\rangle$ isospin states. 

Usually, individual three-body energies are referred to the three-body threshold, which amounts to taking $\xi_\mathrm{core}=0$.
However, when comparing the energy between different three-body systems, a unique zero energy point has to be chosen, which
implies that, when different cores are involved, only the energy of one of them can be taken equal to zero, whereas the energy
of the other ones has to be referred to the first one.

As we can see, the effect of the core hamiltonian is a shift of the effective potentials, different for the halo
state, the IAS, and the AAS, and which is dictated by the core energies.
We have assumed that ${\cal H}_\mathrm{core}$ does not mix the $|t_c,t_c^z=-t_c\rangle$ and $|t_c,t_c^z=-t_c+1\rangle$
core states.

\subsection{Potential matrix elements}

At this point it should be noticed that the isospin wave functions in
Eqs.(\ref{isohalogen}), (\ref{isoIASgen}), and (\ref{isoAASgen})
explicitly show the three-body structures involved in each case, with a core described by the isospin state
$|t_c,t_c^z\rangle$, and a halo state described by $|t_h,t_h^z\rangle$. In particular, the halo state contains a single
configuration with the core and the halo in the $|t_c,t_c^z=-t_c\rangle$ and the $|t_h,t_h^z=-t_h\rangle$ isospin states,
respectively. On the contrary, the IAS and AAS mix two configurations, one with the core and the halo in the
$|t_c,t_c^z=-t_c\rangle$ and $|t_h,t_h^z=-t_h+1\rangle$ isospin states, and the other one with the core and the halo
in the $|t_c,t_c^z=-t_c+1\rangle$ and  $|t_h,t_h^z=-t_h\rangle$ isospin states.

In any case, in order to perform the calculations, 
this decomposition of the isospin states is not the most convenient one. This is because the core-nucleon interactions will in general depend on the total core-nucleon isospin, $t_{cN}$, resulting from the coupling between the core and nucleon isospins.  For this reason, it is preferable to rotate the isospin wave function $|(t_2,t_3)t_h,t_c;T,T_z\rangle$, which is written in what we usually
call the first Jacobi set (or T-set), into the second or third Jacobi sets (Y-sets), where the core and the nucleon isospins are first coupled into $t_{cN}$, and then coupled to the isospin
of the second halo nucleon in order to produce the total isospin $T$ of the system. In this way, the dependence of the isospin wave functions on $t_{cN}$ appears explicitly. In particular,
when rotating into the second Jacobi set we get:
\begin{widetext}
\begin{equation}
|(t_2,t_3)t_h,t_c;T,T_z\rangle=(-1)^{t_c+t_2+t_3}\sum_{t_{cN}} (-1)^{t_{cN}+t_2}
\sqrt{2t_{cN}+1} \sqrt{2t_h+1}
\left\{
\begin{array}{ccc}
t_3 & t_2 & t_h \\
T   &  t_c &  t_{cN}
\end{array}
\right\}
|(t_3,t_c)t_{cN},t_2;T,T_z\rangle,
\label{tcnexp}
\end{equation}
\end{widetext}
which makes evident that the isospin states given in Eqs.(\ref{isohalogen}), (\ref{isoIASgen}), and (\ref{isoAASgen}), can, in general, mix different values of the core-nucleon isospin, $t_{cN}$.  The curly brackets denote the six-$j$ symbol.

When solving the three-body problem, it is always necessary at some
point to compute the matrix elements of the two-body potentials
between the different terms of the basis set used in the expansion of the wave function, 
$\{|{\cal Y}_q\rangle |T T_z\rangle\}$. Thanks to the expression in Eq.(\ref{tcnexp}), and assuming that the core-nucleon
potential, $V_{cN}$, does not mix different $t_{cN}$ values, we easily get for the potential matrix elements:
\begin{widetext}
\begin{equation}
\langle {\cal Y}_q; T,T_z | V_{cN} | {\cal Y}_{q^\prime}; T,T_z \rangle=
\sum_{t_{cN}} (2t_{cN}+1) (2t_h+1) 
\left\{
\begin{array}{ccc}
t_3 & t_2 & t_h \\
T   &  t_c &  t_{cN}
\end{array}
\right\}^2
\langle {\cal Y}_q; (t_3,t_c)t_{cN},t_2;T,T_z |
  V_{cN}^{(t_{cN})}  |{\cal Y}_{q^\prime};(t_3,t_c)t_{cN},t_2;T,T_z\rangle,
\label{vcnmat}
\end{equation}
\end{widetext}
where we can see that the calculation of the different core-nucleon potential matrix elements
requires knowledge of the potentials for the different possible values of $t_{cN}$, i.e., knowledge of 
all the possible $V_{cN}^{(t_{cN})}$ potentials. 

Furthermore, 
in order to fully determine the potential matrix elements, it is also necessary to decouple $t_3$ and $t_c$, 
such that the values of the third isospin components, $t_3^z$ and $t_c^z$, are made explicit. This is required 
in order to know what precise core state and nucleon state are actually interacting.
After doing this we get:
\begin{widetext}
\begin{eqnarray}
\lefteqn{ \hspace*{-2cm}
\langle {\cal Y}_q; (t_3,t_c)t_{cN},t_2;T,T_z |
  V_{cN}^{(t_{cN})}  |{\cal Y}_{q^\prime};(t_3,t_c)t_{cN},t_2;T,T_z\rangle= } \nonumber \\ & & \hspace*{1cm}
  \sum_{t_{cN}^z t_2^z}
  \langle t_{cN} t_2; t_{cN}^z t_2^z|T T_z\rangle^2
  \sum_{t_c^zt_3^z} \langle t_3 t_c; t_3^z t_c^z|t_{cN},t_{cN}^z \rangle^2
  \langle {\cal Y}_q; t_c t_c^z; t_3 t_3^z   | V_{cN}^{(t_{cN})} | {\cal Y}_{q^\prime}; t_c t_c^z; t_3 t_3^z \rangle,
  \label{vt3tc}
\end{eqnarray}
\end{widetext}
which, after insertion into Eq.(\ref{vcnmat}), gives the detail of the required potential matrix elements.  The matrix element, 
$\langle {\cal Y}_q; t_c t_c^z; t_3 t_3^z   | V_{cN}^{(t_{cN})} | {\cal Y}_{q^\prime}; t_c t_c^z; t_3 t_3^z \rangle$,
refers to the matrix element between the basis elements ${\cal Y}_q$ and ${\cal Y}_{q^\prime}$, when the core
is in the $|t_c, t_c^z\rangle$ isospin state, and the nucleon in the $|t_3,t_3^z\rangle$ state, coupled to the two-body
isospin $t_{cN}$, and interacting via the potential $V_{cN}^{(t_{cN})}$. We have assumed here that the potential
does not mix different isospin states.

Therefore, in order to describe a particular system, all the core-nucleon potentials required to compute all the possible
$\langle {\cal Y}_q; t_c t_c^z; t_3 t_3^z   | V_{cN}^{(t_{cN})} | {\cal Y}_{q^\prime}; t_c t_c^z; t_3 t_3^z \rangle$
matrix elements need to be specified. Once this is done, we can then compute the three-body wave function, $\Psi_{3b}$,
entering in Eq.(\ref{wffull}) as described in Eq.(\ref{wf}), which contains the isospin dependence of the potentials 
through the matrix elements in Eqs.(\ref{vcnmat}) and (\ref{vt3tc}), as well as the dependence on the excitation energy of the core through the matrix elements in Eqs.(\ref{coreh}) to (\ref{coreAAS}).

\subsection{Beta decay strength}

Let us now consider the beta decay of some initial state $\Psi^\mathrm{(in)}=\Psi_{3b}^\mathrm{(in)} |T',T'_z\rangle$ 
by the action of the beta decay operator $\hat{\cal O}_{\beta}^A$, as given in Eq.(\ref{eq:decay}), and project
on some final state $\Psi^\mathrm{(fin)}=\Psi_{3b}^\mathrm{(fin)} |T,T_z\rangle$, where $|T',T'_z\rangle$ and $|T,T_z\rangle$
describe the respective initial and final isospin states.

For Fermi and Gamow-Teller decays the corresponding transition strengths are given by:
\begin{equation}
B= \frac{1}{2J_i+1} 
\left| \langle \Psi_{3b}^\mathrm{(fin)}   || \langle T,T_z|
\hat{\cal O}_\beta^A|T',T'_z\rangle || \Psi_{3b}^\mathrm{(in)} \rangle \right|^2, 
\label{fstr}
\end{equation}
where $\hat{\cal O}_\beta^A$ is either the Fermi,
${\cal O}_F^\mp= \sum_j t_j^{\pm}$, or Gamow-Teller, ${\cal O}_{GT}^{\mp} = \sum_j \bm{\sigma}_j t_j^{\pm}$, decay 
operators, and where $J_i$ is the angular momentum quantum number of the initial state and $j$ runs over all the $A$ nucleons of the initial system.

As shown in Eq.(\ref{eq:decay}), the Fermi and Gamow-Teller operators can be split in two pieces, the first one involving 
the decay of the valence nucleons and the second one involving the decay of the core nucleons. In this way,
the reduced matrix elements contained in Eq.(\ref{fstr}) can be written as: 
\begin{eqnarray}
\lefteqn{\hspace*{-1cm}
\langle \Psi_{3b}^\mathrm{(fin)}  
 ||\langle T,T_z| \hat{\cal O}_\beta^A |T',T'_z\rangle || \Psi_{3b}^\mathrm{(in)} 
 \rangle = }  \nonumber \\ &&
\langle \Psi_{3b}^\mathrm{(fin)}
 ||\langle T,T_z| \hat{\cal O}_\beta^{A_h} |T',T'_z\rangle || \Psi_{3b}^\mathrm{(in)}
 \rangle  \nonumber \\ && +
\langle \Psi_{3b}^\mathrm{(fin)}
 ||\langle T,T_z| \hat{\cal O}_\beta^{A_c}|T',T'_z\rangle || 
    \Psi_{3b}^\mathrm{(in)} \rangle, 
    \label{eq16}
\end{eqnarray}
where 
\begin{equation}
\hat{\cal O}_\beta^{A_h}=\sum_{j=v_1,v_2} t_j^\pm \mbox{ or } 
\hat{\cal O}_\beta^{A_h}=\sum_{j=v_1,v_2}  \bm{\sigma}_j t_j^\pm
\end{equation}
describes the Fermi or Gamow-Teller decay of the halo
nucleons, denoted by $v_1$ and $v_2$, and 
\begin{equation}
\hat{\cal O}_\beta^{A_c}=\sum_{j \in \mathrm{core}}  t_j^\pm \mbox{ or } 
\hat{\cal O}_\beta^{A_c}=\sum_{j \in \mathrm{core}}  \bm{\sigma}_j t_j^\pm
\end{equation}
describes the Fermi or Gamow-Teller decay of the $A_c$ nucleons in the core.

From Eqs.(\ref{isohalogen}), (\ref{isoIASgen}), and (\ref{isoAASgen}), it is simple to see that the core isospin matrix element 
in the equation above becomes:
 \begin{eqnarray}
 \lefteqn{\hspace*{-1.5cm}
 \langle \mathrm{halo}; T,T_z| \hat{\cal O}_\beta^{A_c}|\mathrm{halo}; T',T'_z\rangle =} \nonumber \\ &&
  \langle t_c,t_c^z=-t_c| \hat{\cal O}_\beta^{A_c}|t'_c,t_c^{\prime z}=-t'_c\rangle,
  \label{ffin1}
 \end{eqnarray}
 \begin{eqnarray}
 \lefteqn{\hspace*{-5mm}
 \langle \mathrm{IAS}; T,T_z| \hat{\cal O}_\beta^{A_c}|\mathrm{halo}; T',T'_z\rangle = } \label{ffin2} \\ &&
 \sqrt{\frac{t_c}{t_c+t_h}}
  \langle t_c,t_c^z=-t_c+1| \hat{\cal O}_\beta^{A_c}|t'_c,t_c^{\prime z}=-t'_c\rangle,
  \nonumber
 \end{eqnarray}
 or
 \begin{eqnarray}
 \lefteqn{\hspace*{-5mm}
 \langle \mathrm{AAS}; T,T_z| \hat{\cal O}_\beta^{A_c}|\mathrm{halo}; T',T'_z\rangle = }  \label{ffin3} \\ &&
  -\sqrt{\frac{t_h}{t_c+t_h}}
  \langle t_c,t_c^z=-t_c+1| \hat{\cal O}_\beta^{A_c}|t'_c,t_c^{\prime z}=-t'_c\rangle, \nonumber
 \end{eqnarray}
which permit to write the reduced beta decay matrix element Eq.(\ref{eq16}) as:
\begin{eqnarray}
\lefteqn{\hspace*{-1cm}
\langle \Psi_{3b}^\mathrm{(fin)}  
 ||\langle T, T_z| \hat{\cal O}_\beta^A|T',T'_z\rangle || 
 \Psi_{3b}^\mathrm{(in)}  \rangle  = }
 \label{eq4}  \\ & &
\langle \Psi_{3b}^\mathrm{(fin)}  ||
 \langle T,T_z| \hat{\cal O}_\beta^{A_h}|T',T'_z\rangle  || \Psi_{3b}^\mathrm{(in)} \rangle 
  \nonumber \\ & &   +  
   f_{\mathrm{fin}} \langle \Psi_{3b}^\mathrm{(fin)}|| \Psi_{3b}^\mathrm{(in)} \rangle 
\langle t_c,t_c^z | \hat{\cal O}_\beta^{A_c} |  t'_c,t^{\prime z}_c \rangle, \nonumber
\end{eqnarray}
where $f_\mathrm{fin}=1,\sqrt{t_c/(t_c+t_h)}$, or $-\sqrt{t_h/(t_c+t_h)}$, depending on the character (halo state, IAS, or AAS) 
of the final state, and where $\langle \Psi_{3b}^\mathrm{(fin)}|| \Psi_{3b}^\mathrm{(in)} \rangle$ is the overlap
of the initial and final three-body wave functions, excluding the isospin part. The final matrix element describes
the beta decay of the core, whose initial and final isospin states are given by $|t'_c,t^{\prime z}_c\rangle$
and $|t_c,t^z_c\rangle$, respectively.

For Fermi decay, the transition operator depends on the isospin only. It is then possible to factorize the
matrix element $\langle \Psi_{3b}^\mathrm{(fin)}  ||
 \langle T,T_z| \hat{\cal O}_\beta^{A_h} |T',T'_z\rangle  || \Psi_{3b}^\mathrm{(in)} \rangle$
into one term involving only the coordinate and spin parts
of the initial and final three-body wave functions ($\Psi_{3b}$ in Eq.(\ref{wffull})), and another matrix element involving only the isospin part. In other words, for Fermi decay we can write:
\begin{eqnarray}
\lefteqn{
\langle \Psi_{3b}^\mathrm{(fin)} ||  \langle T,T_z|
  \sum_{j=1}^{A}  t_j^+ |T,T'_z\rangle || \Psi_{3b}^\mathrm{(in)} \rangle  = 
\langle \Psi_{3b}^\mathrm{(fin)}  || \Psi_{3b}^\mathrm{(in)}  \rangle  }  \label{eq6} \\ && \hspace*{-3mm} \times
\left(\langle T,T_z|  \sum_{j=v_1,v_2}   t_j^+ |T',T'_z\rangle  + f_\mathrm{fin}
\langle  t_c,t_c^z | \sum_{j \in \mathrm{core} }
 t_j^+ |  t'_c,t^{\prime z}_c \rangle 
\right),
\nonumber
\end{eqnarray}
which means that the Fermi decay matrix element 
is given just by the overlap of the initial and final three-body wave functions (excluding 
the isospin part) times an isospin dependent global factor.

\subsection{Contribution from the continuum background}

The matrix elements derived in the previous section correspond to the beta decay transition from some
halo initial state into some either halo, or IAS, or AAS, final states, each of them characterized by
specific isospin quantum numbers. However, when computing the strength for a particular transition, it is necessary 
to include the contribution from transitions into the continuum states having the same quantum numbers as
the final state, either halo state, IAS, or AAS.

In this work the continuum states will be computed as discrete states after imposing a box boundary
condition. In other words, for a given set of quantum numbers, the radial wave functions in Eq.(\ref{wf}) are obtained
by solving the radial part of the three-body equations in a box with a sufficiently maximum large value for the hyperradius.
To be specific, we shall use the finite-difference method described for instance in Chapter 19 of Ref.~\cite{pre97}. This
method permits to reduce the system of coupled differential equations into an eigenvalue problem, such that 
the eigenfunctions are the radial solutions evaluated at specific chosen values of the radial coordinate, and the 
eigenvalues are the “discrete” continuum energies.

In this way, together with the precise three-body wave function for the final state (halo state, IAS, or AAS),
we get, for each of them, a family of
``discrete'' continuum states with the same quantum numbers. Therefore, the total transition strength
to the states with a definite set of quantum numbers will be given by the sum of all the strengths 
given in Eq.(\ref{fstr})
 for Fermi or Gamow-Teller strengths, respectively. The sum runs of course over all the computed
discrete states.

At this point it is also interesting to consider the differential transition strength, which permits to obtain
the distribution of the strength as a function of the final three-body energy. When dealing with discrete states,
and following Eq.(\ref{fstr}), we can easily write the differential transition strength for Fermi and 
Gamow-Teller transitions as:
\begin{eqnarray}
\lefteqn{
\frac{dB}{dE}= \frac{1}{2J_i+1} \times } \label{dstr} \\ && \hspace*{-4mm}
\sum_k \delta(E-E_k)
\left| \langle \Psi_{3b}^\mathrm{(fin)}(E_k)   || \langle T,T_z|
 \hat{\cal O}_\beta^A |T',T'_z\rangle || \Psi_{3b}^\mathrm{(in)} \rangle \right|^2, 
 \nonumber
\end{eqnarray}
where $k$ runs over all the discrete final states, with energy $E_k$, and wave function 
$\Psi_{3b}^{\mathrm{(fin)}}(E_k)$. The integral over the three-body energy, $E$, trivially provides the
total strength, and the reduced matrix elements for the Fermi and Gamow-Teller decays are given by 
Eq.(\ref{eq4}), which for Fermi decay can be simplified into Eq.(\ref{eq6}).

In our calculations the $\delta$-function will be replaced by the normalized Gaussian
\begin{equation}
\delta(E-E_k) \approx \frac{1}{\sqrt{2\pi}\sigma} e^{-\frac{(E-E_k)^2}{2\sigma^2}},
\label{eq10}
\end{equation}
where the Gaussian width, $\sigma$, is made as small as possible, but producing a smooth strength 
function (otherwise Eq.(\ref{dstr}) would become a sequence of delta functions located at the energies of the 
different discrete states). Typical $\sigma$ values range within the interval
$\sigma=0.1$ MeV and $\sigma=0.3$ MeV and therefore larger than the
distance between the discrete continuum states, which, in our calculations, will be in the range of
0.05 MeV to 0.1 MeV.

\section{Application to $^{11}$Li beta decay}

We shall here employ the extended three-body formalism described in
the previous section to investigate beta-decay of the $^{11}$Li ground
state.  After specifying the isospin formalism in the context, we
proceed to calculate energies, potentials and wave functions.

\subsection{Specific information}

In the following we denote the ground state wave function by
$^{11}$Li$_{\mathrm{g.s.}}$, which decays into different $^{11}$Be
states.  The $^{11}$Li$_{\mathrm{g.s.}}$ state will be constructed as
a $^9$Li core in its ground state, $^9$Li$_{\mathrm{g.s.}}$, and two
halo neutrons.

For $^{11}$Be we shall consider as well the case of a two-neutron halo system having $^9$Be in its ground state,
 $^9$Be$_{\mathrm{g.s.}}$, as core. We shall denote this state as $^{11}$Be$_{\mathrm{g.s.}}$. Together with it,
 we shall consider as well transitions into the IAS of $^{11}$Li$_{\mathrm{g.s.}}$, denoted as $^{11}$Be$_{\mathrm{IAS}}$, and into the AAS of $^{11}$Li$_{\mathrm{g.s.}}$, denoted as $^{11}$Be$_{\mathrm{AAS}}$. 
 
It is important to note again that beta decay does not change the spatial structure of the system. Therefore, all
the three $^{11}$Be states considered here,  $^{11}$Be$_{\mathrm{g.s.}}$, $^{11}$Be$_{\mathrm{IAS}}$,
and $^{11}$Be$_{\mathrm{AAS}}$, are assumed to have the same partial wave structure ($s$, $p$, and $d$ partial wave components) as
$^{11}$Li$_{\mathrm{g.s.}}$, and they all have spin and parity $3/2^-$ (so, contrary to $^{11}$Li$_{\mathrm{g.s.}}$,
$^{11}$Be$_{\mathrm{g.s.}}$ is not 
the ground state of $^{11}$Be, and the subscript refers only to having $^9$Be in its ground state as core).

It should be noted from the outset that our calculations will not at
the moment be able to describe all of the decay. The decay of the
core, $^{9}$Li$_{\mathrm{g.s.}}$, proceeds about half of the time to
$^9$Be$_{\mathrm{g.s.}}$, and the rest of the time to excited states
that are typically thought of as $^8$Be+n and $^5$He+$\alpha$
combinations before decaying into a final $\alpha\alpha$n continuum. Three-body calculations of the resonances in $^{9}$Be are doable \cite{Alv10}, but extending them to five-body calculations by adding two nucleons is beyond current capabilities.

Since beta decay does not change the spatial wave function, the main strength will, if we neglect the small change in Coulomb energy, go to states with similar energy, and for the Fermi strength even the same energy. For $^{11}$Li this implies that most strength should lie within about 5 MeV of the IAS as the Gamow-Teller Giant Resonance, that includes most of the GT strength, is roughly at the IAS energy in this region of the  nuclear chart \cite{Sag93}. This is an excitation energy where many channels are open and continuum degrees of freedom are therefore important.

For the halo decay, the GT sum rule gives a total strength of 6. The
$^{11}$Li halo consists mainly of $s_{1/2}^2$ and $p_{1/2}^2$
configurations. In a halo decay one of these neutrons is transformed
into a proton; in most cases the ensuing configuration will not differ
much in energy (and could be treated in few-body models), the main
exception being for a final $p_{3/2}$ proton that could combine with
the $^{9}$Li core to produce states in $^{10}$Be. The remaining
$p_{1/2}$ neutron may then combine to form low-lying states in
$^{11}$Be, a prime example being the first excited $1/2^-$ state
\cite{Suz94}. Beta transitions of this type (moving a halo particle
into the core) cannot easily be reproduced in our few-body calculations.

For core decays, transitions that involve the $^{9}$Be$_\mathrm{IAS}$ are straightforward to treat theoretically, even though it is inaccessible in the $^{9}$Li decay. For all other states fed in $^{9}$Be one has to treat them individually, find their interactions with the halo neutrons and extract the relevant $^{9}$Li to $^{9}$Be$^*$ matrix elements either from experiment or from other theoretical models. We shall attempt this for the $^{9}$Be ground state, but will not be able to treat other states. The consequences of this will be discussed in section \ref{sec:comparison}, as will the extent to which the $^{11}$Li strength resembles the $^{9}$Li strength.

\subsection{The isospin wave functions} 

From Eq.(\ref{isohalogen}) we can get the isospin wave functions for
$^{11}$Li$_{\mathrm{g.s.}}$ and $^{11}$Be$_{\mathrm{g.s.}}$, which are given by:

\begin{eqnarray}
|\mbox{$^{11}$Li}_{\mathrm{g.s.}} \rangle&=&
|(t_2,t_3)t_h=1, t_c=\frac{3}{2} ; T=\frac{5}{2},T_z=-\frac{5}{2}\rangle \nonumber \\ &=&
|t_c= \frac{3}{2}t_c^z=-\frac{3}{2}\rangle |t_h=1,t_h^z=-1\rangle  \nonumber \\ 
&\equiv& |^9\mbox{Li}_{\mathrm{g.s.}}+n+n\rangle,
\label{iso11Li}
\end{eqnarray}
and 
\begin{eqnarray}
|\mbox{$^{11}$Be}_{\mathrm{g.s.}} \rangle&=&|(t_2,t_3)t_h=1, t_c=\frac{1}{2} ; 
T=\frac{3}{2},T_z=-\frac{3}{2}\rangle   \nonumber \\ &=&
|t_c=\frac{1}{2}t_c^z=-\frac{1}{2}\rangle |t_h=1,t_h^z=-1\rangle  \nonumber \\
&\equiv& |^9\mbox{Be}_{\mathrm{g.s.}}+n+n\rangle,
\label{gs11Be}
\end{eqnarray}
where the core states $|t_c=\frac{3}{2},t_c^z=-\frac{3}{2}\rangle$ and $|t_c=\frac{1}{2},t_c^z=-\frac{1}{2}\rangle$
correspond to $^9$Li$_{\mathrm{g.s.}}$  and $^9$Be$_{\mathrm{g.s.}}$, respectively. Therefore, 
$^{11}$Li$_{\mathrm{g.s.}}$ and $^{11}$Be$_{\mathrm{g.s.}}$
correspond to three-body structures having either
$^9$Li$_{\mathrm{g.s.}}$ or $^9$Be$_{\mathrm{g.s.}}$ as core, surrounded by two neutrons
(since $t_h^z=-1$, the halo is necessarily formed by two neutrons).

Similarly, for $^{11}$Be$_{\mathrm{IAS}}$ and $^{11}$Be$_{\mathrm{AAS}}$, making use of Eq.(\ref{isoIASgen})
and Eq.(\ref{isoAASgen}), we get:
\begin{widetext}
\begin{eqnarray}
|\mbox{$^{11}$Be}_{\mathrm{IAS}} \rangle&=&|(t_2,t_3)t_h=1,t_c=\frac{3}{2} ; T=\frac{5}{2},T_z=-\frac{3}{2}\rangle \nonumber \\  &=&
\sqrt{\frac{2}{5}}  |t_c=\frac{3}{2}t_c^z=-\frac{3}{2}\rangle |t_h=1,t_h^z=0\rangle +\sqrt{\frac{3}{5}}  |t_c=\frac{3}{2}t_c^z=-\frac{1}{2}\rangle |t_h=1,t_h^z=-1\rangle
 \nonumber \\ 
&\equiv&   \sqrt{\frac{2}{5}}  |\mbox{$^9$Li$_{\mathrm{g.s.}}$}+n+p\rangle+ \sqrt{\frac{3}{5}}|\mbox{$^9$Be}_{\mathrm{IAS}}+n+n\rangle
\label{isoIAS} 
\end{eqnarray}
and
\begin{eqnarray}
|\mbox{$^{11}$Be}_{\mathrm{AAS}} \rangle&=&|(t_2,t_3)t_h=1,t_c=\frac{3}{2}; T=\frac{3}{2},T_z=-\frac{3}{2}\rangle \nonumber \\  &=&
\sqrt{\frac{3}{5}}  |t_c=\frac{3}{2}t_c^z=-\frac{3}{2}\rangle |t_h=1,t_h^z=0\rangle -\sqrt{\frac{2}{5}}  |t_c=\frac{3}{2}t_c^z=-\frac{1}{2}\rangle |t_h=1,t_h^z=-1\rangle
 \nonumber \\ 
&\equiv&  \sqrt{\frac{3}{5}}  |\mbox{$^9$Li$_{\mathrm{g.s.}}$}+n+p\rangle - \sqrt{\frac{2}{5}}|\mbox{$^9$Be}_{\mathrm{IAS}}+n+n\rangle,
\label{isoAAS}
\end{eqnarray}
\end{widetext}
where, as before, $|t_c=\frac{3}{2},t_c^z=-\frac{3}{2}\rangle$ corresponds to $^9$Li$_{\mathrm{g.s.}}$, but $|t_c=\frac{3}{2},t_c^z=-\frac{1}{2}\rangle$ corresponds to $^9$Be  populating the IAS of $^9$Li$_\mathrm{g.s.}$,
 i.e., $^9$Be$_{\mathrm{IAS}}$.
Therefore, both, $^{11}$Be$_{\mathrm{IAS}}$ and $^{11}$Be$_{\mathrm{AAS}}$, mix two different three-body structures, one having $^9$Li$_{\mathrm{g.s.}}$ 
 as core, plus one neutron and one proton (as it corresponds to $t_h^z=0$),
and a second system having $^9$Be$_{\mathrm{IAS}}$ as core, plus two neutrons. As expected, the weights are such that the IAS and AAS described in Eqs.(\ref{isoIAS})
and (\ref{isoAAS}) are orthogonal.

Note that for all the three-body systems considered here, $^{11}$Li$_{\mathrm{g.s.}}$, $^{11}$Be$_{\mathrm{g.s.}}$,
$^{11}$Be$_{\mathrm{IAS}}$, and $^{11}$Be$_{\mathrm{AAS}}$, the two halo nucleons couple to $t_h=1$. 
This means that the isospin function is always symmetric under exchange of both nucleons, and therefore
the three-body wave function describing the relative motion between the three constituents, $\Psi_{3b}$, has
to be in all the cases antisymmetric under the same exchange.

\subsection{Potential matrix elements}
\label{potmatel}

As mentioned, in order to obtain the three-body wave functions, it is always necessary, at some point, to 
compute the matrix elements $\langle {\cal Y}_q; T,T_z |V_{cN}^{(t_{cN})}  |{\cal Y}_{q^\prime}; T,T_z\rangle$, 
Eq.(\ref{vcnmat}), which involve the different core-nucleon interactions and the basis terms formed by the 
usual hyperspherical harmonics, ${\cal Y}_q$, and the isospin function of the system, $|T T_z\rangle$. 
These matrix elements are, in general, computed numerically, and, as shown in Eq.(\ref{vt3tc}), they permit to determine 
the two-body potentials needed in order to perform the calculation.

The particularization of the potential matrix elements for the systems considered in this work, 
$^{11}$Li$_{\mathrm{g.s.}}$, $^{11}$Be$_{\mathrm{g.s.}}$, $^{11}$Be$_{\mathrm{IAS}}$, and $^{11}$Be$_{\mathrm{AAS}}$,
can be made by use of the isospin wave functions given in Eqs.(\ref{iso11Li}) to (\ref{isoAAS}). The details
are given in Appendix~\ref{app1}.

In particular, from Eqs.(\ref{9lin}) to (\ref{aasnp}) we get that, in order to obtain the three-body wave functions
for the $^{11}$Li$_{\mathrm{g.s.}}$, $^{11}$Be$_{\mathrm{g.s.}}$,
$^{11}$Be$_{\mathrm{IAS}}$, and $^{11}$Be$_{\mathrm{AAS}}$ states, the following core-nucleon interactions need
to be specified: $i)$ The interaction between
the $^9$Li$_{\mathrm{g.s.}}$ core and the neutron (only $t_{cN}=2$ is possible), $ii)$ the interaction
between the $^9$Be$_{\mathrm{g.s.}}$ core and the neutron (only $t_{cN}=1$ is possible),
$iii)$ the interaction between the $^9$Li$_{\mathrm{g.s.}}$ core and the proton for $t_{cN}=2$,
$iv)$ the interaction between the $^9$Be$_{\mathrm{IAS}}$ core and the neutron for $t_{cN}=2$,
$v)$ the interaction between the $^9$Li$_{\mathrm{g.s.}}$ core and the proton for $t_{cN}=1$,
and $vi)$ the interaction between the $^9$Be$_{\mathrm{IAS}}$ core and the neutron for $t_{cN}=1$.

\subsection{Core energy}

In section \ref{coren} we described how the core hamiltonian, ${\cal H}_{\mathrm{core}}$, gives rise to a
shift of the effective three-body potentials determining the energy of the different three-body
states. This shift is governed by the energy of the core, which, as we can see from Eqs.(\ref{iso11Li})
to (\ref{isoAAS}), for the systems considered here can be either $^{9}$Li$_{\mathrm{g.s.}}$,
$^{9}$Be$_{\mathrm{g.s.}}$, or $^{9}$Be$_{\mathrm{IAS}}$, whose respective energies
will be denoted by $\xi_{\mbox{\scriptsize $^9$Li$_{\mathrm{g.s.}}$}}$,
$\xi_{\mbox{\scriptsize $^9$Be$_{\mathrm{g.s.}}$}}$,
 and $\xi_{\mbox{\scriptsize $^9$Be}_{\mathrm{IAS}}}$.
 
The details of the calculation of these core matrix elements are given in Appendix~\ref{app2},
and, more specifically, the shift of the effective three-body potentials is given by Eqs.(\ref{matcor0})
to (\ref{matcor2}).

\subsection{Relative energies}

\begin{figure*}[t]
\centering
\includegraphics[width=15cm]{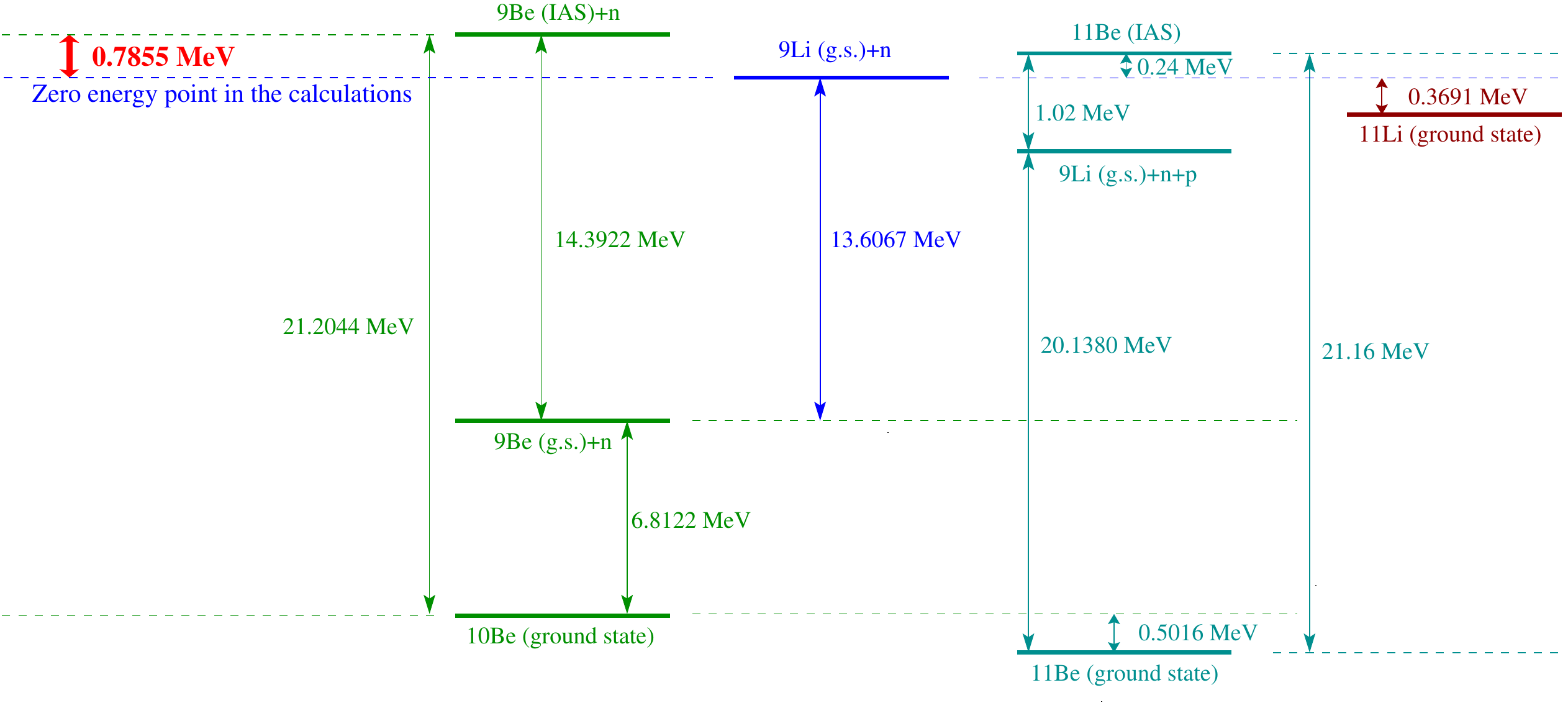}
\caption{Relative energies between the different states involved in the calculations. Data from Ref.~\cite{til04}. 
To keep the figure clean, the separation between the states is not always properly scaled.}
\label{levels}
\end{figure*}

In this work we are describing four different three-body systems, i.e., $^{11}$Li$_{\mathrm{g.s.}}$,
$^{11}$Be$_{\mathrm{g.s.}}$,  $^{11}$Be$_{\mathrm{IAS}}$, and $^{11}$Be$_{\mathrm{AAS}}$. When
computed individually, it is common to refer the energy of the three-body states to three-body thresholds. In 
other words, the zero-energy point is taken as the energy of the core for each of them.

However, in order to describe the decay of some system ($^{11}$Li$_{\mathrm{g.s.}}$ in our case)
into some other systems (three different $^{11}$Be states in our calculations) it is necessary to refer
all the energies to a single zero-energy value. In our calculations we have chosen the zero-energy point 
as the $^{11}$Li$_\mathrm{g.s.}$ threshold, which implies that 
$\xi_{\mbox{\scriptsize $^9$Li$_{\mathrm{g.s.}}$}}=0$ in Eq.(\ref{core1}).
After this choice, as one can find in \cite{til04}, we have  that 
$\xi_{\mbox{\scriptsize $^9$Be}_{\mathrm{g.s.}}}=-13.6067$ MeV
and $\xi_{\mbox{\scriptsize $^9$Be}_{\mathrm{IAS}}}=0.7855$ MeV.

The relative energies between the different systems involved in our calculations are sketched in  Fig.~\ref{levels} 
(data from  Ref.~\cite{til04}). The values of $\xi_{\mbox{\scriptsize $^9$Be}_{\mathrm{g.s.}}}$
and $\xi_{\mbox{\scriptsize $^9$Be}_{\mathrm{IAS}}}$ given above amount to an excitation energy
of 14.3922 MeV of $^9$Be$_{\mathrm{IAS}}$ with respect to $^9$Be$_{\mathrm{g.s.}}$,
which in turn is 6.8122 MeV above the $^{10}$Be ground state energy. This leads to
 an energy threshold of 21.2044 MeV for the $^9$Be$_{\mathrm{IAS}}$+neutron state above the $^{10}$Be ground state. 
 Also, the experimental excitation energy of $^{11}$Be$_{\mathrm{IAS}}$ with respect to
 $^{11}$Be$_{\mathrm{g.s.}}$, 21.16 MeV, gives rise to an excitation energy of only 0.24 MeV with 
 respect to the $^{11}$Li$_{\mathrm{g.s.}}$ threshold.

\section{Three-body calculations}

In order to perform the three-body calculations, the key quantities are the two-body interactions 
between the three constituents of the systems. We now describe how these interactions are constructed, 
and discuss the three-body wave functions obtained after the subsequent calculations.

\subsection{Two-body potentials}
\label{2bodypot}

The systems involved in this work are $^{11}$Li$_{\mathrm{g.s.}}$, $^{11}$Be$_{\mathrm{g.s.}}$, $^{11}$Be$_{\mathrm{IAS}}$, or $^{11}$Be$_{\mathrm{AAS}}$. The required core-nucleon interactions
are the six interactions mentioned in section~\ref{potmatel}, which, together with the nucleon-nucleon potential,
constitute all the interactions needed for the calculations.

For the nucleon-nucleon interaction we use the potential given in Eq.(5) of Ref.~\cite{gar04}. This is a simple 
potential employing Gaussian shapes for the central, spin-spin, spin-orbit, and tensor terms, and whose parameters
(strength and range) are adjusted to reproduce the experimental $s$- and $p$-wave nucleon-nucleon scattering lengths
and effective ranges.
 
\begin{table}
\begin{center}
\begin{tabular}{c|ccc}
   &  $t_{cN}=2$   & $t_{cN}=1$  &$n-^9$Be$_{\mathrm{g.s.}}$ \\ \hline
 Pots.  &  $i), iii), iv)$ & $v), vi)$  &  $ii)$  \\
$S_c^{(s)}$        &  $-5.4$    & $-23.61$ &   $-11.47$ \\
$S_c^{(p)}$        &  $260.75$  & 246.70 &   $-1.89$ \\ 
$S_c^{(d)}$        &  $260   $  & 246.70    &  $-44.22$    \\ \hline
$S_{ss}^{(s)}$    &   $-4.5$    & $-1.83$ &   $0.59$ \\
$S_{ss}^{(p)}$    &   $1.0 $   &  $-3.35$&   $-4.0$ \\ 
$S_{ss}^{(d)}$    &   $-9.0 $   & $-3.35$   &  $-0.88$        \\ \hline
$S_{so}^{(p)}$    &   300     &  300   &     30.0 \\
$S_{so}^{(d)}$    &   $-300$  & $-300$  & $-10.0$         \\
\end{tabular}
\end{center}
\caption{Strengths, in MeV, of the nuclear Gaussian potentials ($V^{(\ell)}(r)=S^{(\ell)} e^{-r^2/b^2}$) 
for the six potentials, labeled as in section~\ref{potmatel}, entering in this work.
The range of the Gaussians is $b=3.0$ fm for potential $ii)$ and $b=2.55$ fm for the other five potentials.}
\label{tab1}
\end{table} 
 
 For all the necessary core-nucleon interactions we take an $\ell$-dependent potential with the form:
 \begin{equation}
 V_{cN}^{(\ell)}(r)=V_c^{(\ell)}(r)+V_{ss}^{(\ell)}(r) \bm{s}_c\cdot \bm{j}_N+V_{so}^{(\ell)}(r) \bm{\ell}\cdot \bm{s}_N,
 \label{potcN}
 \end{equation}
 with central, spin-spin, and spin-orbit terms (plus Coulomb interaction when needed). In the expression above $\bm{s}_c$ and $\bm{s}_N$ are the spins of the core and the nucleon,
 respectively, $\bm{\ell}$ is the relative orbital angular momentum between them, and $\bm{j}_N$ results from the coupling of $\bm{\ell}$ and $\bm{s}_N$. This choice of the spin-spin 
 and spin-orbit operators is particularly useful to preserve the shell-model quantum numbers for the nucleon state, making easier the implementation of the Pauli principle \cite{gar03}. In all the cases the potential functions
$V_c^{(\ell)}$, $V_{ss}^{(\ell)}$, and $V_{so}^{(\ell)}$ are taken to be Gaussians whose parameters are adjusted to 
reproduce the available experimental information about the core-nucleon system.

The connection to an explicit isospin dependence is described in Appendix~\ref{app3}, where the expression of the potential can be found for use if the isospin formalism is preferred.

Below we describe how the different potentials, $i)$, $ii$), $iii)$, $iv)$, $v)$, and $vi)$, as labeled 
in section~\ref{potmatel}, are selected:

$i)$ and $iii)$ For the neutron-$^9$Li$_{\mathrm{g.s.}}$ interaction (potential $i)$), which enters in the matrix
elements (\ref{9lin}), (\ref{iasnp}), and (\ref{aasnp}), we use the interaction specified in Ref.~\cite{Gar20} to describe the $^{11}$Li ground state. The given Gaussian
potentials are shallow enough to avoid binding of a halo nucleon into a Pauli forbidden state. In particular,
since the $p_{3/2}$-shell is fully occupied by the neutrons in the core, the $p$-wave spin-orbit potential
is used to push the $p_{3/2}$-states away. As shown in Eq.(\ref{rot1}),
 the  neutron-$^9$Li$_{\mathrm{g.s.}}$ interaction corresponds fully to a core-neutron isospin $t_{cN}=2$. 
 The precise parameters of the potential for $s$, $p$, and $d$ waves are given in the second column of Table~\ref{tab1}.  
 The same interaction is used for the proton-$^9$Li$_{\mathrm{g.s.}}$ potential with $t_{cN}=2$ (potential $iii)$), 
 entering in the matrix elements (\ref{iasnp}) and (\ref{aasnp}), simply by adding the corresponding 
 core-proton Coulomb repulsion.
 
$ii)$ For the neutron-$^9$Be$_{\mathrm{g.s.}}$ interaction the potential is fitted to reproduce the known $^{10}$Be 
properties \cite{til04}. The $^9$Be core has spin and parity $J^\pi=\frac{3}{2}^-$, which, when coupled to 
an $s_{1/2}$-neutron, gives rise to a $1^-$ and a $2^-$ state. Experimentally, $^{10}$Be is known to have
such a 1$^-$ and $2^-$ states, both bound with respect to the $^9$Be-neutron threshold by 0.8523 MeV and 0.5489 MeV, respectively. These two states are then used to construct the $s$-wave potential. In the same way the $2^+$ state, bound by 0.8538 MeV, is used to obtain the $p$-wave 
potential. Experimentally, there is no low-lying $1^+$ state, and the spin-spin interaction is then used to push away
this state. Also, since the partial wave structure of the system is the same as the one in 
$^{11}$Li$_{\mathrm{g.s.}}$, the spin-orbit potential is employed to move away the Pauli forbidden
$p_{3/2}$ states. Finally, the $4^-$ and $3^-$ states, with excitation energies 9.27 MeV and 10.15 MeV,
respectively, are used to estimate the potential parameter for $d$-waves.
This interaction enters in the matrix element (\ref{9ben}) only, and it corresponds to $t_{cN}=1$. The parameters of the
neutron-$^9$Be$_{\mathrm{g.s.}}$ potential are given in the fourth column of Table~\ref{tab1}.
 
$iv)$ For the neutron-$^9$Be$_{\mathrm{IAS}}$ interaction and $t_{cN}=2$ we take into account that
$^{9}$Be$_{\mathrm{IAS}}$  is the IAS of the $^9$Li$_{\mathrm{g.s.}}$.
The members of a given isobaric multiplet preserve the nuclear structure, in such a way that the energy 
difference between them is provided almost exclusively by the different strength of the Coulomb repulsion. 
For this reason, when $t_{cN}=2$, we use for the  neutron-$^9$Be$_{\mathrm{IAS}}$ interaction,
which enters in the matrix elements (\ref{iasnp}) and (\ref{aasnp}), the same nuclear potential as
for the neutron-$^9$Li$_{\mathrm{g.s.}}$ interaction, which is given in the second column 
of Table~\ref{tab1}.

$v)$ and $vi)$ For the proton-$^9$Li$_{\mathrm{g.s.}}$ and neutron-$^9$Be$_{\mathrm{IAS}}$
potentials with $t_{cN}=1$, since, as mentioned above, the $^9$Li$_{\mathrm{g.s.}}$ 
and $^9$Be$_{\mathrm{IAS}}$ cores belong to the same isospin multiplet, we take in both cases the 
same nuclear potential. The Coulomb repulsion is the only difference between both interactions.

For the case of the $^9$Be$_{\mathrm{IAS}}$+neutron system, the $t_{cN}=1$ states are the Anti-Analog States of the
$^{10}$Li low-lying states, whose energy is about 20 MeV above the $^{10}$Be ground state, as
shown in Fig.~\ref{levels}. Similar pairs of known IAS-AAS partners suggest that the Anti-Analog States lie a few 
MeV below the Isobaric Analog States. Therefore, the $^9$Be$_{\mathrm{IAS}}$+neutron ($t_{cN}=1$) states are expected
to be located about 16 to 18 MeV above the $^{10}$Be ground state. Good candidates are therefore the experimentally known
$^{10}$Be states with excitation energies 17.12 MeV, 17.79 MeV, and 18.55 MeV \cite{til04}. This is actually
supported by the fact that, for the $^9$Be$_{\mathrm{IAS}}$+proton ($^{10}$B) system, which except for the Coulomb
repulsion is equivalent to the $^9$Be$_{\mathrm{IAS}}$+neutron system, there is a known series of 
isospin 1 states at an energy of about 18 MeV above the $^{10}$Be ground state, among which a $2^-$, a $2^+$, and a $1^-$ state have been observed.

For this reason we have assumed
 that the first and third of the mentioned $^{10}$Be states, at 17.12 MeV and 18.55 MeV, are $2^-$ and $1^-$ states arising from the coupling of an $s_{1/2}$ neutron and the core ($J^\pi=\frac{3}{2}^-$), and the second
state, at 17.79 MeV, is assumed to be a $2^+$ state coming from the coupling of a $p_{1/2}$ neutron and the core. As in 
the neutron-$^9$Li$_{\mathrm{g.s.}}$ interaction, the $p_{3/2}$ neutron states are pushed away.
The precise parameters are given in the third column of Table~\ref{tab1}. For the $d$-waves we use the
same potential as for the $p$-waves but changing the sign in the spin-orbit potential (as systematically happens
for the other potentials).
At this point it is important to keep in mind the relative energies between all the systems involved in our calculations,
 which are shown in Fig.~\ref{levels}. The $2^-$, $2^+$, and $1^-$ states in $^{10}$Be  used to determine the $t_{cN}=1$ part of the interaction, with excitation energies all around 18 MeV above the $^{10}$Be ground state, are actually below the $^9$Be$_{\mathrm{IAS}}$-neutron threshold, and therefore they are bound with respect to this threshold   
by $4.08$ MeV, $3.41$ MeV, and $2.65$ MeV, respectively.

  An additional simplification comes from treating
the $^9$Be$_{\mathrm{IAS}}$ core as a single particle;
$^9$Be$_{\mathrm{IAS}}$ is far from being bound, although it is quite
narrow (less than one keV). One could envisage describing the
$^9$Be$_{\mathrm{IAS}}$+neutron system as a four-body
($\alpha+\alpha+n+n$) system, and consequently the corresponding
$^{11}$Be states as a five-body system. In any case this is out of our
reach, as it would make the calculations significantly more complex.

\subsection{Three-body wave functions}

 Using the interactions as described above, we have made the three-body calculations for $^{11}$Li$_{\mathrm{g.s.}}$,
 $^{11}$Be$_{\mathrm{g.s.}}$, $^{11}$Be$_{\mathrm{IAS}}$, and $^{11}$Be$_{\mathrm{AAS}}$. Among them the
 more tricky ones are the calculations of $^{11}$Be$_{\mathrm{IAS}}$ and $^{11}$Be$_{\mathrm{AAS}}$,
 since, as shown in Eqs.(\ref{iasnp}) and (\ref{aasnp}), they both combine different core-nucleon potentials, and,
 in the case of $^{11}$Be$_{\mathrm{AAS}}$, also different values of $t_{cN}$.
 
 \begin{figure}
\centering
\includegraphics[width=\linewidth]{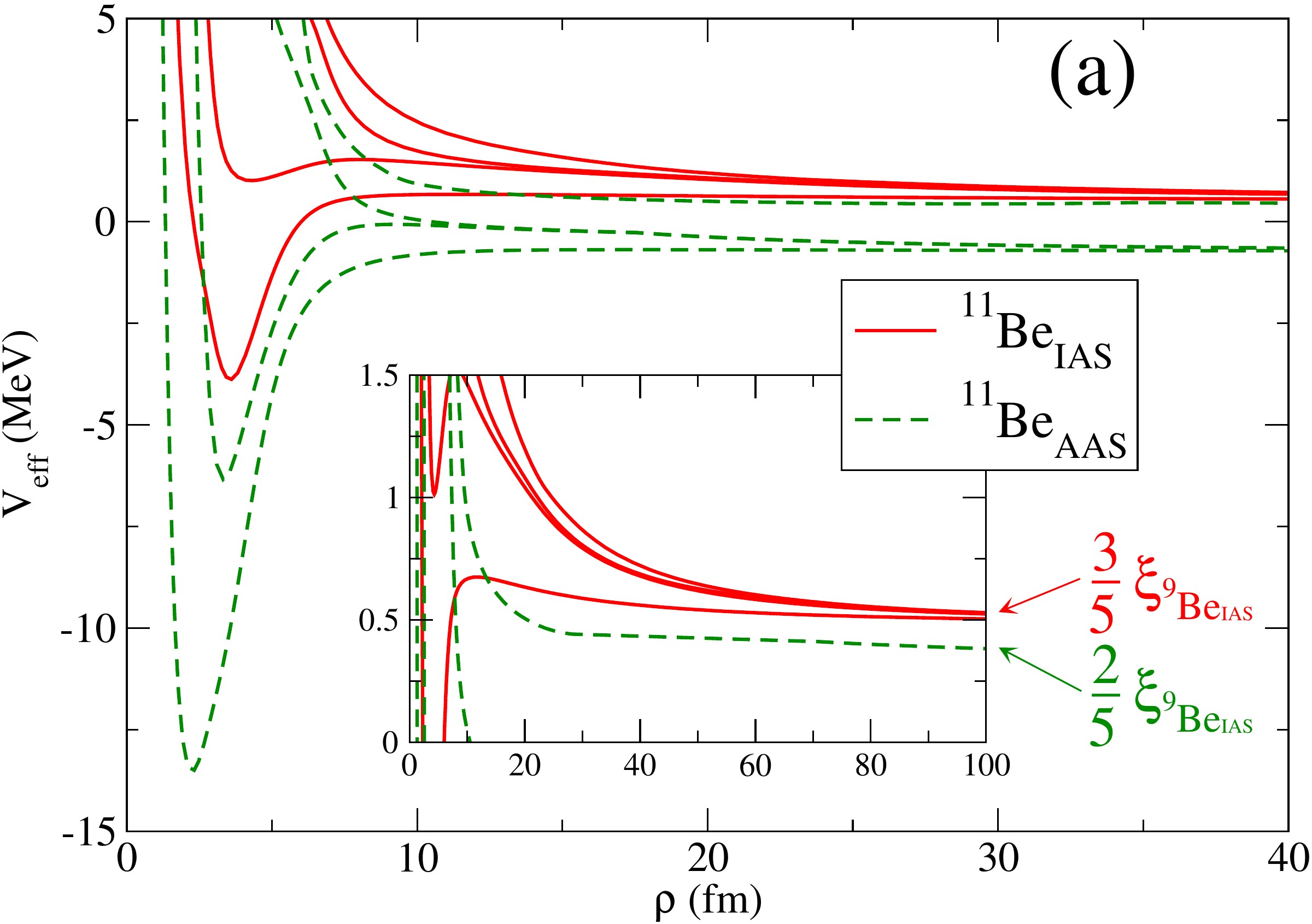}
\includegraphics[width=\linewidth]{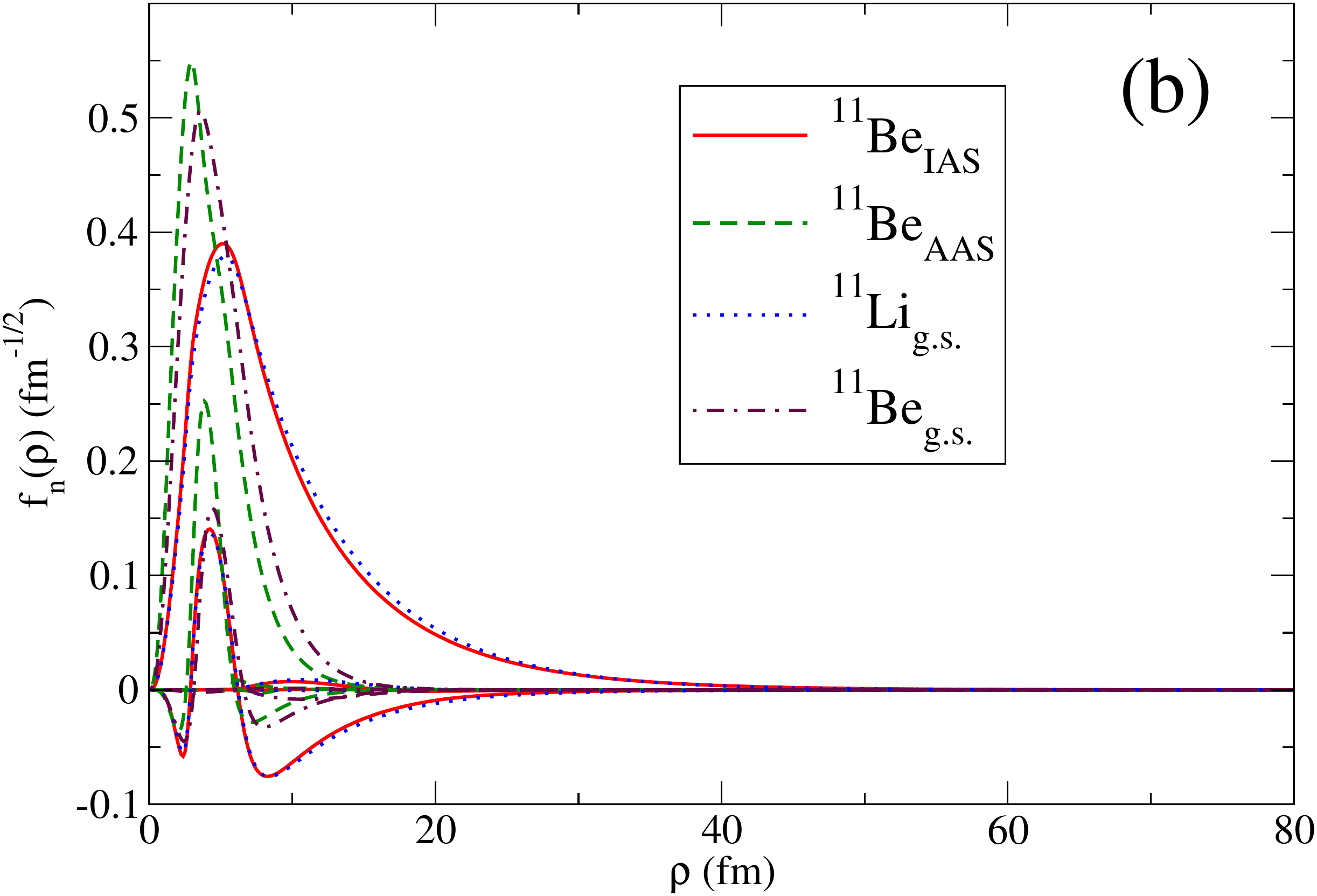}
\caption{(a) The four lowest effective potentials for the IAS (solid red) and AAS (dashed green) in $^{11}$Be.The inset shows a zoom of the large distance behavior.  (b) Radial wave functions for the four lowest adiabatic terms for 
$^{11}$Be$_{\mathrm{IAS}}$ (solid red), $^{11}$Be$_{\mathrm{AAS}}$ (dashed green), $^{11}$Li$_{\mathrm{g.s.}}$
(dotted blue), and $^{11}$Be$_{\mathrm{g.s.}}$ (dot-dashed brown).}
\label{figpots}
\end{figure}
 
 In Fig.~\ref{figpots}a we show for these two systems the four lowest computed effective adiabatic potentials 
 as defined in Eq.(\ref{veff}).
 The solid-red and dashed-green curves correspond, respectively, to the $^{11}$Be$_{\mathrm{IAS}}$
 and the $^{11}$Be$_{\mathrm{AAS}}$ systems. We can immediately see that the 
 potentials for the AAS are clearly deeper than the ones for the IAS. This is expected, since in the AAS case the interaction contains the $t_{cN}=1$ part of the potential, 
 which, as explained above, binds by several MeV the lowest $2^-$, $1^-$, and $2^+$ states in the $^9$Be$_{\mathrm{IAS}}$-neutron system with respect to the $^9$Be$_{\mathrm{IAS}}$-neutron threshold. In fact, the lowest AAS potentials go 
 asymptotically to the energies of the bound $^9$Be$_{\mathrm{IAS}}$-neutron states, although weighted with the geometric factors involved in the matrix element in Eq.(\ref{matpot2}).
 
 The higher potentials in the AAS case, as well as the ones for the IAS, go asymptotically to the value determined by the energy shift imposed by the core 
 hamiltonian in Eqs.(\ref{matcor1}) and (\ref{matcor2}), i.e., $\frac{3}{5} \xi_{\mbox{\scriptsize $^9$Be}_{\mathrm{IAS}}}$ and $\frac{2}{5} \xi_{\mbox{\scriptsize $^9$Be}_{\mathrm{IAS}}}$ for the IAS and AAS, respectively,
 where  $\xi_{\mbox{\scriptsize $^9$Be}_{\mathrm{IAS}}}=0.7855$ MeV (see Fig.~\ref{levels}). This is seen in the inset in Fig.~\ref{figpots}a, which shows the large distance part of the potentials.

 As shown in Fig.~\ref{levels}, the experimental excitation energy of $^{11}$Be$_{\mathrm{IAS}}$ is 21.16 MeV, which is 
 0.24 MeV above the $^9$Li$_{\mathrm{g.s.}}$+$n$+$n$ threshold that 
 we take as our zero-energy point. This value of the IAS energy, 0.24 MeV, is below the asymptotic energy limit, $\frac{3}{5}\xi_{\mbox{\scriptsize $^9$Be}_{\mathrm{IAS}}}=0.47$ MeV, of the potentials shown in the inner panel in Fig.~\ref{figpots}a, which implies that the computed radial wave function will decay exponentially at large distances, as a regular bound state. The same of course happens for the AAS, which, as one can anticipate from Fig.~\ref{figpots}a, is going to be clearly more bound than the IAS.

In the calculation for $^{11}$Li$_{\mathrm{g.s.}}$ and $^{11}$Be$_{\mathrm{IAS}}$ it has been necessary to introduce a 
small attractive effective three-body force in such a way that the respective experimental energies, 
$-0.369$ MeV and 0.24 MeV with respect to the energy threshold shown in Fig.~\ref{levels}, are well reproduced.
For $^{11}$Be$_{\mathrm{AAS}}$ the same effective three-body force as for the IAS has been used, and a 
binding energy of $-5.53$ MeV with respect to our zero energy point has been obtained. This amounts to an excitation 
energy of this $^{11}$Be state of 15.39 MeV, which is consistent with observations in other IAS/AAS partners.
Also, the computed energy difference between the IAS and AAS is equal to 5.77 MeV, consistent as well 
 with the upper limit of 7 MeV estimated in Appendix~\ref{app3}.

Finally, for $^{11}$Be$_{\mathrm{g.s.}}$ the three-body calculation without any three-body potential
gives rise to a binding energy
of $-18.03$ MeV with respect to our chosen zero-energy value. This means, see Fig.~\ref{levels}, that the computed 
$^{11}$Be$_{\mathrm{g.s.}}$ wave function has an excitation energy of 2.89 MeV with respect to the 
ground state of $^{11}$Be.
Since this excitation energy is in the vicinity of the experimentally known 3/2$^-$ states in $^{11}$Be
with excitation energy 2.69 MeV and 3.97 MeV, we then have not included any three-body
potential for $^{11}$Be$_{\mathrm{g.s.}}$.

\begin{table}
\begin{tabular}{l|cc}
                             & Exc. energy &  $^9$Li$_{\mathrm{g.s.}}$+$n$+$n$  \\ \hline
$^{11}$Li$_{\mathrm{g.s.}}$  &  0  & $-0.37$ \\
$^{11}$Be$_{\mathrm{g.s.}}$  &  2.89 & $-18.03$ \\
$^{11}$Be$_{\mathrm{IAS}}$   &  21.16  &  0.24\\
$^{11}$Be$_{\mathrm{AAS}}$   &  15.39 &  $-5.53$
\end{tabular}
\caption{Computed excitation energies (second column) and energies relative to the $^9$Li$_{\mathrm{g.s.}}$+$n$+$n$ threshold
(third column) for $^{11}$Li$_{\mathrm{g.s.}}$, $^{11}$Be$_{\mathrm{g.s.}}$, $^{11}$Be$_{\mathrm{IAS}}$, and $^{11}$Be$_{\mathrm{AAS}}$. All energies are given in MeV. }
\label{taben}
\end{table}

The computed energies for $^{11}$Li$_{\mathrm{g.s.}}$,
 $^{11}$Be$_{\mathrm{g.s.}}$, $^{11}$Be$_{\mathrm{IAS}}$, and $^{11}$Be$_{\mathrm{AAS}}$ are summarized
 in Table~\ref{taben}, where the second column gives the excitation energies and the third one the same energies
but relative to the $^9$Li$_{\mathrm{g.s.}}$+$n$+$n$ threshold used in this work. All the energies are given in MeV.

In Fig.~\ref{figpots}b we plot the radial wave functions associated to the four lowest adiabatic terms, Eq.(\ref{wf}), for 
each of the systems computed in this work, i.e., 
$^{11}$Be$_{\mathrm{IAS}}$ (solid red), $^{11}$Be$_{\mathrm{AAS}}$ (dashed green), $^{11}$Li$_{\mathrm{g.s.}}$
(dotted blue), and $^{11}$Be$_{\mathrm{g.s.}}$ (dot-dashed brown). As we can see, in all the cases the first two terms provide almost the full wave function. The wave functions for $^{11}$Be$_{\mathrm{IAS}}$ and $^{11}$Li$_{\mathrm{g.s.}}$
are very similar. This is due to the fact that in both cases only the $t_{CN}=2$ part of the core-nucleon potential enters.
Since the cores in both systems belong to the same isospin multiplet the nuclear interaction is therefore the same for the
two of them. The only difference 
comes from the Coulomb repulsion present in the $^9$Li$_{\mathrm{g.s.}}$+$n$+$p$ structure entering in the IAS case, see Eq.(\ref{isoIAS}).

\section{Beta decay strength}

The energy distribution for a beta decay transition is given by Eq.(\ref{dstr}).
The initial state is $^{11}$Li$_{\mathrm{g.s.}}$, whose isospin wave function 
is given by Eq.(\ref{iso11Li}). For the final states we consider either $^{11}$Be$_{\mathrm{g.s.}}$,
Eq.(\ref{gs11Be}), $^{11}$Be$_{\mathrm{IAS}}$, Eq.(\ref{isoIAS}), or $^{11}$Be$_{\mathrm{AAS}}$, Eq.(\ref{isoAAS}).

\subsection{Overlap between initial and final wave functions.} One of the ingredients that enter in both, Fermi
and Gamow-Teller, transition strengths is the overlap between the initial and final three-body wave functions 
(excluding the isospin part), $\Psi_{3b}^\mathrm{(in)}$ and $\Psi_{3b}^\mathrm{(fin)}$. In order to investigate this overlap, we introduce the overlap function:

\begin{equation}
F_\mathrm{ov}(\rho)=\int \rho^5 \Psi_{3b}^\mathrm{(fin)}(\rho,\Omega) \Psi_{3b}^\mathrm{(in)}(\rho,\Omega) d\Omega,
\label{frhof}
\end{equation}
where the integral is over the five hyperangles only. With this definition we have that:
\begin{equation}
\int d\rho F_\mathrm{ov}(\rho) = \langle \Psi_{3b}^\mathrm{(fin)} | 
\Psi_{3b}^\mathrm{(in)} \rangle = 
\frac{1}{\sqrt{2J_i +1}} \langle \Psi_{3b}^\mathrm{(fin)} || 
\Psi_{3b}^\mathrm{(in)} \rangle. \label{intfov}
\end{equation}

\begin{figure}
\centering
\includegraphics[width=\linewidth]{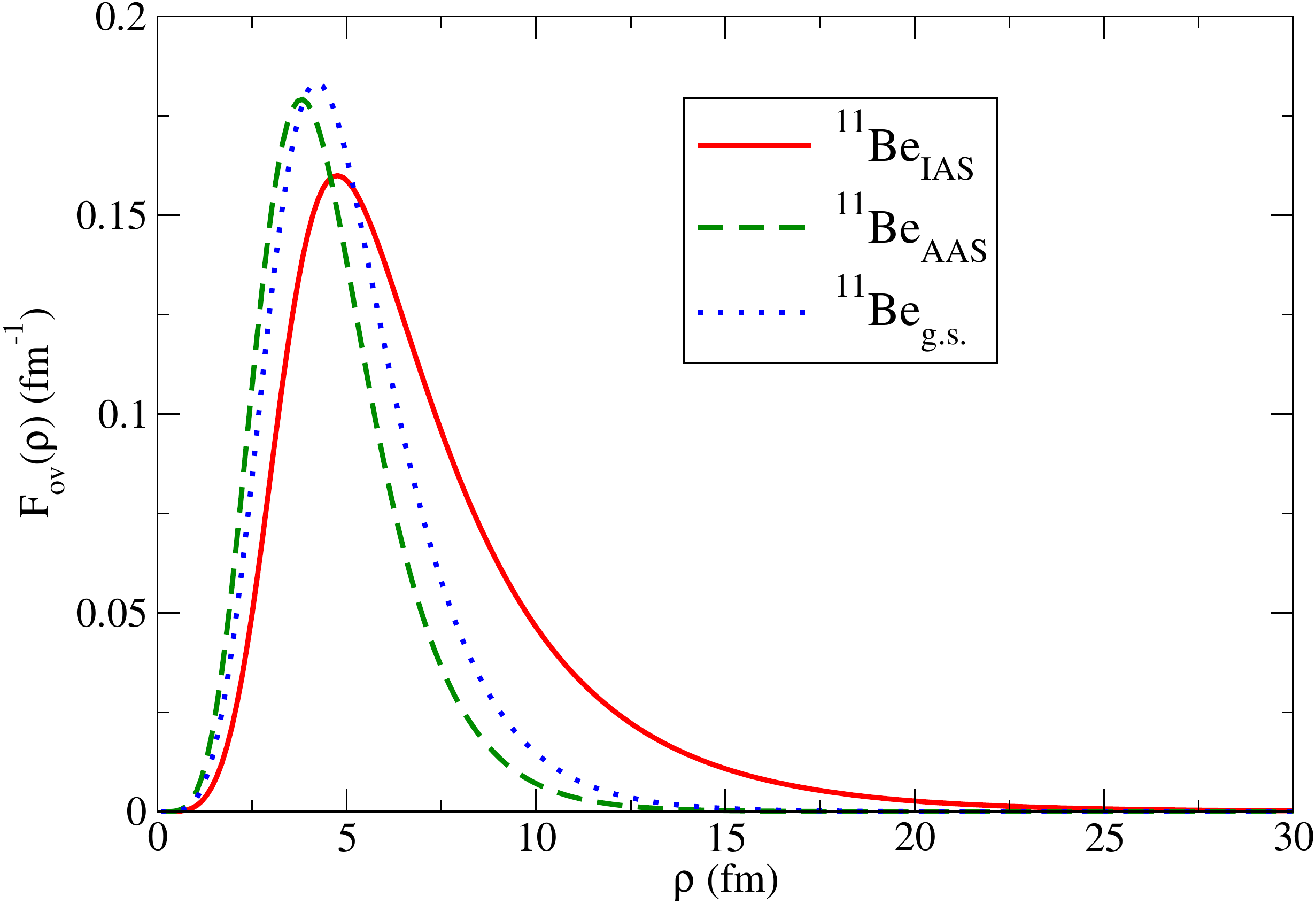}
\caption{$F_\mathrm{ov}(\rho)$ function in Eq.(\ref{frhof}) for the overlap between the $^{11}$Li$_{\mathrm{g.s.}}$ wave 
function and the one corresponding to $^{11}$Be$_{\mathrm{IAS}}$ (solid), $^{11}$Be$_{\mathrm{AAS}}$ (dashed),
and $^{11}$Be$_{\mathrm{g.s.}}$ (dotted).}
\label{figov}
\end{figure}

In Fig.~\ref{figov} we show the overlap function in Eq.(\ref{frhof}) where $\Psi_{3b}^\mathrm{(in)}$ is the wave function
of $^{11}$Li$_{\mathrm{g.s.}}$ and $\Psi_{3b}^\mathrm{(fin)}$ is the wave function
of either $^{11}$Be$_{\mathrm{IAS}}$ (solid curve), $^{11}$Be$_{\mathrm{AAS}}$ (dashed curve), or
$^{11}$Be$_{\mathrm{g.s.}}$ (dotted curve). Due to the similarity between the $^{11}$Li$_{\mathrm{g.s.}}$ 
and $^{11}$Be$_{\mathrm{IAS}}$ wave functions, dotted 
and solid curves in Fig.~\ref{figpots}b, the integral over $\rho$, Eq.(\ref{intfov}), of the corresponding overlap function (solid curve in 
Fig.~{\ref{figov}}) is 0.99, very close to 1.
As we can also anticipate from Fig.~\ref{figpots}b, the cases of $^{11}$Be$_{\mathrm{AAS}}$ and
$^{11}$Be$_{\mathrm{g.s.}}$ are different. These wave functions are clearly narrower than the one of 
$^{11}$Li$_{\mathrm{g.s.}}$, and, as a consequence, the integral over $\rho$ of the overlap function
(dashed and dotted curves in  Fig.~{\ref{figov}}) will be significantly smaller, 0.72 and 0.82 
for $^{11}$Be$_{\mathrm{AAS}}$ and$^{11}$Li$_{\mathrm{g.s.}}$, respectively. 

\begin{figure}
\centering
\includegraphics[width=\linewidth]{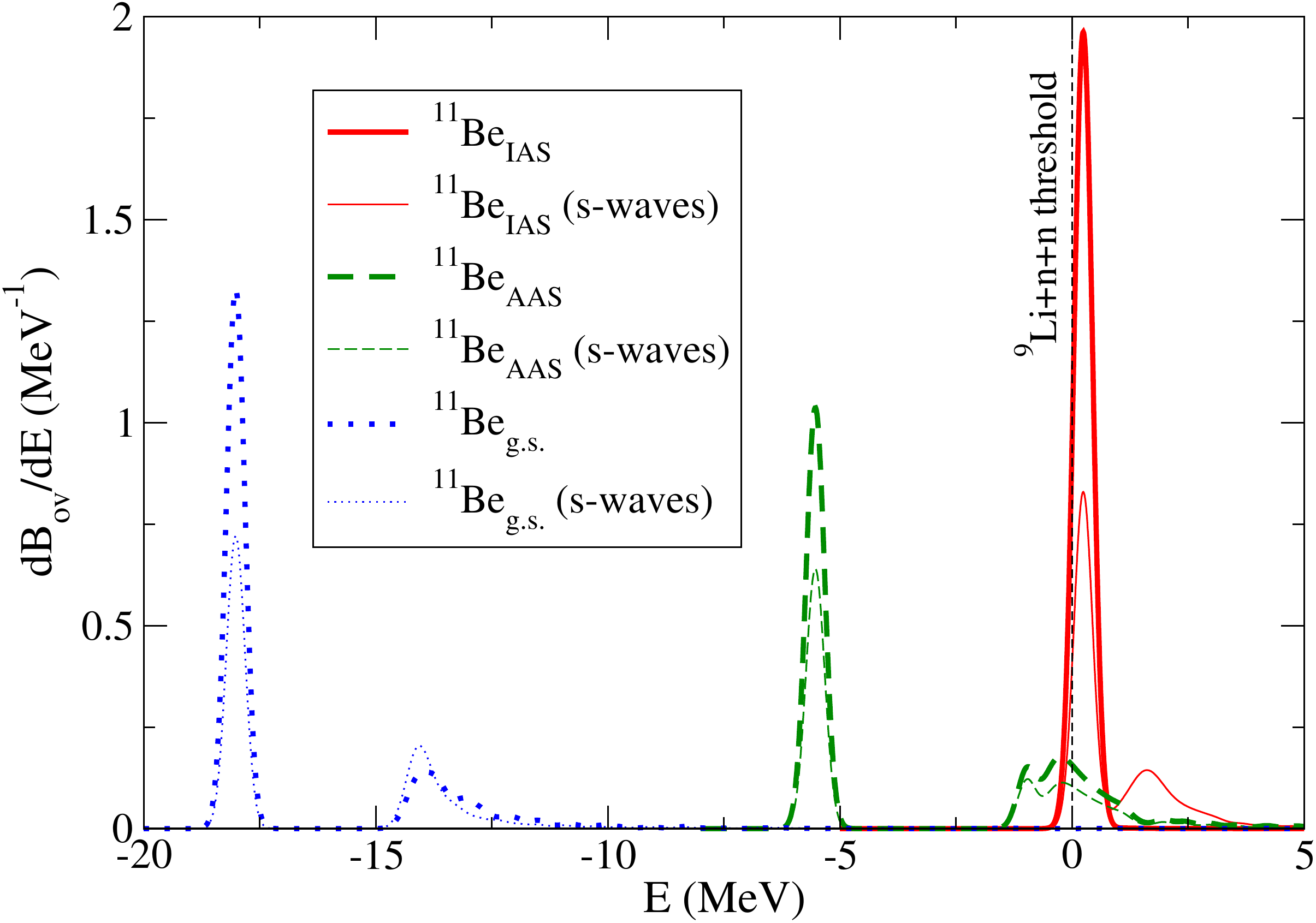}
\caption{Overlap strength, as given in Eq.(\ref{dfov}), as a function of the three-body energy, 
for transitions into states, bound and continuum, with the same quantum numbers as $^{11}$Be$_{\mathrm{IAS}}$ (solid-red), $^{11}$Be$_{\mathrm{AAS}}$ (dashed-green),
and $^{11}$Be$_{\mathrm{g.s.}}$ (dotted-blue). The thin curves show the contribution from core-nucleon $s$-waves only.}
\label{fig1}
\end{figure}

At this point it is interesting to introduce the effect of the continuum background in the overlap functions
by means of the differential distribution:
\begin{equation}
\frac{dB_\mathrm{ov}}{dE}=\frac{1}{2J_i+1} 
\sum_k \delta(E-E_k)
\left| \langle \Psi_{3b}^\mathrm{(fin)}(E_k)  || \Psi_{3b}^\mathrm{(in)} \rangle \right|^2, 
 \label{dfov}
\end{equation}
which is analogous to the differential transition strength introduced in Eq.(\ref{dstr}),
and where $k$ runs over all the discretized continuum states. 

The result from Eq.(\ref{dfov})
is shown in Fig.~\ref{fig1} for transitions into states, bound and continuum, with the same quantum numbers, as 
$^{11}$Be$_{\mathrm{IAS}}$ (solid-red), $^{11}$Be$_{\mathrm{AAS}}$ (dashed-green), and 
$^{11}$Be$_{\mathrm{g.s.}}$ (dotted-blue). The thin curves in the figure 
give the contribution when only the relative core-nucleon $s$-waves are considered.
The three main peaks are produced by the transitions into the specific
$^{11}$Be$_\mathrm{IAS}$, $^{11}$Be$_\mathrm{AAS}$, and $^{11}$Be$_\mathrm{g.s.}$ states, which 
provide most of the contribution to $B_\mathrm{ov}$. The remainder is the contribution
from the continuum states. We can see there is a non-negligible contribution from the continuum states
for transitions into the states with the $^{11}$Be$_\mathrm{AAS}$ and $^{11}$Be$_\mathrm{g.s.}$ quantum
numbers. For the $^{11}$Be$_\mathrm{IAS}$, the $s$-wave contribution (thin-solid) has a bump at about 2 MeV,
but this bump disappears from the total due to the interference with the other partial waves, in such a
way that the contribution from the continuum states is negligible.

\begin{table}
\begin{tabular}{l|ccc}
    &   $^{11}$Be$_\mathrm{IAS}$ & $^{11}$Be$_\mathrm{AAS}$ &  $^{11}$Be$_\mathrm{g.s.}$ \\ \hline
$B_\mathrm{ov}$ (total) & 0.99  &  0.87   & 0.90 \\ \hline
$B_\mathrm{ov}$ (no cont.)  &  $0.98 (99\%)$ & 0.52 (60\%)   & 0.67 (74\%) \\
$B_\mathrm{ov}$ (continuum) &  $0.01 (1\%) $ & 0.35 (40\%)   & 0.23 (26\%) \\ \hline
$B_\mathrm{ov}$ ($s$-waves) & 0.63 (64\%)  & 0.56 (64\%)   & 0.59 (66\%) \\
$B_\mathrm{ov}$ ($p$-waves) & 0.30 (30\%) & 0.28 (32\%) &  0.23  (26\%)  \\
$B_\mathrm{ov}$ ($d$-waves) & 0.00 (0\%)  & 0.00 (0\%) &   0.00 (0\%)    \\ 
\end{tabular}
\caption{Integrated overlap function, $B_\mathrm{ov}$, from Eq.(\ref{dfov}), for the states with
same quantum numbers as $^{11}$Be$_\mathrm{IAS}$, $^{11}$Be$_\mathrm{AAS}$, and $^{11}$Be$_\mathrm{g.s.}$.
The following two rows give the contributions to the total from the non-continuum states and the continuum background, respectively. The last three rows give the contribution to the total when only relative core-nucleon $s$-waves, $p$-waves, 
and $d$-waves are included. 
The values within parenthesis give the percentage over the total given by each contribution.}
\label{tab2}
\end{table} 

The integral over $E$ of the different curves in Fig.~\ref{fig1} gives the total value, $B_\mathrm{ov}$,
of the overlap, which is 0.99, 0.87, and 0.90 for states with the same quantum numbers as 
$^{11}$Be$_\mathrm{IAS}$, $^{11}$Be$_\mathrm{AAS}$, and $^{11}$Be$_\mathrm{g.s.}$, respectively. 
The total contribution from the specific $^{11}$Be$_\mathrm{IAS}$, $^{11}$Be$_\mathrm{AAS}$, and 
$^{11}$Be$_\mathrm{g.s.}$ states is given by the square of Eq.(\ref{intfov}), which, as obtained
after the analysis of Fig.~\ref{figov}, is equal to 0.98, 0.52, and 0.67 for the three cases.
In other words, the contribution from the continuum background is very small ($\sim 0.01$) for
the IAS case, but quite relevant, 0.35 and 0.23, for the other two cases. These results
are summarized in the first three rows in Table~\ref{tab2}. 

In the table, in the last three rows, we also give the contribution to the total when only the relative 
core-nucleon $s$-waves, $p$-waves, and $d$-waves are included. As we can see, the contribution to the
overlaps from the $d$-waves is negligible in all the three cases. Also, the weight of $s$ and $p$
partial waves is similar in the three cases too, especially if we consider the percentage over the total
given by each of them, which is indicated in the table within parenthesis. Note that the sum of the $s$-,
$p$-, and $d$-contributions does not provide the total value of $B_\mathrm{ov}$. The missing
contribution, which ranges from 4\% to 8\% depending on the case, is the contribution from the $s$-, 
$p$-, and $d$-interferences when squaring
the overlap in Eq.(\ref{intfov}), which, by construction, are omitted when the calculation is limited
to $s$-, $p$-, or $d$-waves.

Note as well that, although the value of the integral over $E$ of Eq.(\ref{dfov}) is independent of
$\sigma$, see Eq.(\ref{eq10}), the height of the peaks of the curves in Fig.~\ref{fig1} is very
sensitive to its value. In this calculation we have taken $\sigma=0.2$ MeV, but a value slightly smaller
can produce curves with clearly higher peaks, but that one could still consider smooth. Therefore,
whereas the curves in Fig.~\ref{fig1} can be seen as an estimate of the differential overlap, the integral 
under the curves, the integrated strengths, Table~\ref{tab2}, which are the relevant quantities in order to obtain the total
transition strengths, are well established results of the calculations.

\subsection{Fermi decay.}
For Fermi transitions, the beta decay matrix element is given by Eq.(\ref{eq6}), which simply is the overlap
$ \langle \Psi_{3b}^\mathrm{(fin)} || \Psi_{3b}^\mathrm{(in)} \rangle$
as given in Eq.(\ref{intfov}), multiplied by a global isospin dependent factor. Therefore, as seen from Eq.(\ref{dstr}), the corresponding
differential transition strength will be the overlap strength defined in Eq.(\ref{dfov}), multiplied
by the isospin global factor squared. In other words, the differential Fermi transition strength for the different
transitions considered in this work, is given by the curves in Fig.~\ref{fig1}, but, each of them, has to be multiplied by the 
corresponding isospin factor squared.

As we can see in Eq.(\ref{eq6}), the isospin factor contains three ingredients, $f_{\mathrm{fin}}$, the halo isospin matrix
element $\langle T,T_z|  \sum_{j=v_1,v_2}   t_j^+ |T',T'_z\rangle$, and the core isospin matrix element 
$\langle  t_c,t_c^z | \sum_{j \in  \mbox{ \scriptsize $^9$Li$_{\mathrm{g.s.}}$}}  t_j^+ |  t'_c=\frac{3}{2},t^{\prime z}_c=-\frac{3}{2} \rangle$.

\begin{table}
\begin{tabular}{c|cccc}
 Daughter  &     $\langle \sum_{j=v_1,v_2}   t_j^+ \rangle$  &   $f_{\mathrm{fin}}$  &
 $\langle  \sum_{j \in  \mbox{ \scriptsize $^9$Li$_{\mathrm{g.s.}}$}}  t_j^+ \rangle$  & Total \\ \hline
$^{11}$Be$_{\mathrm{g.s.}}$ & 0                     &          1             &  0 &  0  \\
$^{11}$Be$_{\mathrm{AAS}}$  &  $\sqrt{\frac{6}{5}}$ & $-\sqrt{\frac{2}{5}}$  &  $\sqrt{3}$ &  0 \\ 
$^{11}$Be$_{\mathrm{IAS}}$  & $\frac{2}{\sqrt{5}}$  & $ \sqrt{\frac{3}{5}}$   &  $\sqrt{3}$ &  $\sqrt{5}$ \\
\end{tabular}
\caption{For decays into $^{11}$Be$_{\mathrm{g.s.}}$, $^{11}$Be$_{\mathrm{AAS}}$, and $^{11}$Be$_{\mathrm{IAS}}$, we
give the values of the halo isospin matrix element (second column), $f_{\mathrm{fin}}$ factor (third column), and
core isospin matrix element (fourth column), entering into the total isospin term (fifth column) in Eq.(\ref{eq6}).}
\label{tobis}
\end{table}

For the decay of $^{11}$Li$_{\mathrm{g.s.}}$ into $^{11}$Be$_{\mathrm{g.s.}}$, due to the different initial and 
final core isospins, $\frac{3}{2}$ in the case of $^{9}$Li$_{\mathrm{g.s.}}$ and
$\frac{1}{2}$ for $^{9}$Be$_{\mathrm{g.s.}}$, both the halo isospin matrix element 
and the core isospin matrix element (which in this case corresponds to $t_c=\frac{1}{2}$ and $t_c^z=-\frac{1}{2}$) are
equal to zero. Therefore
the isospin factor is in this case equal to zero, and there is no Fermi transition between $^{11}$Li$_{\mathrm{g.s.}}$ 
and $^{11}$Be$_{\mathrm{g.s.}}$. These values are collected into the second row of Table~\ref{tobis}.

For transitions from $^{11}$Li$_{\mathrm{g.s.}}$ into $^{11}$Be$_{\mathrm{AAS}}$, as shown in Eq.(\ref{ffin3})
for $t_c=3/2$ and $t_h=1$, we have that $f_{\mathrm{fin}}=-\sqrt{2/5}$. In this case the halo isospin matrix element takes the value
$\sqrt{6/5}$, and the core isospin matrix element, with $t_c=\frac{3}{2}$ and $t_c^z=-\frac{1}{2}$, is equal to $\sqrt{3}$.
When these values are combined to get the isospin factor in Eq.(\ref{eq6}), we obtain that its value is equal to zero
(third row of Table~\ref{tobis}).
The Fermi transition between $^{11}$Li$_{\mathrm{g.s.}}$ and $^{11}$Be$_{\mathrm{AAS}}$ is forbidden as it connects states with different $T$.

Finally, for transitions from $^{11}$Li$_{\mathrm{g.s.}}$ into $^{11}$Be$_{\mathrm{IAS}}$, Eq.(\ref{ffin2})
for $t_c=3/2$ and $t_h=1$ results into $f_{\mathrm{fin}}=\sqrt{3/5}$. In this case the halo isospin matrix element is equal to 
$2/\sqrt{5}$, and, as for transitions into the AAS, the core isospin matrix element is $\sqrt{3}$. All this results
in a global isospin term in Eq.(\ref{eq6}) equal to $\sqrt{5}$ (again, these values are collected
in Table~\ref{tobis}, last row). Therefore, a Fermi transition between 
$^{11}$Li$_{\mathrm{g.s.}}$ and $^{11}$Be$_{\mathrm{IAS}}$ is possible, and the corresponding differential
transition strength is given by the solid (red) curve in Fig.~\ref{fig1}, but multiplied by a factor of 5.
As a consequence, as $B_\mathrm{ov}=0.99$ in this case (see Table~\ref{tab2}), the total Fermi strength is also
approximately equal to 5 for this transition.

As mentioned, the products of a beta decay process have, in general, a non well-defined isospin. This quantum
number is not a fundamental symmetry, since the long-range Coulomb interaction breaks the symmetry.
Note that isospin breaking has been found to be small in earlier explicit calculations for halo states \cite{Suz91,Han93}.

Therefore, the daughter state of the $^{11}$Li$_\mathrm{g.s.}$ nucleus could actually mix both, the $^{11}$Be$_{\mathrm{IAS}}$
and the $^{11}$Be$_{\mathrm{AAS}}$ states.
In order to estimate the isospin admixture of these states we closely follow the procedure described
in \cite{Han93} (more precisely, Eqs.(7) and (8) of that reference). The matrix element between the spatial
wave functions of the IAS and AAS of the Coulomb potential, $Z_c e^2/r$ (where $Z_c$ is the $^9$Li charge,
$e^2=1.44$ MeV$\times$fm, and $r$ is the core-proton distance) is equal to 1.31 MeV. For the Coulomb displacement
of the core, $\Delta E_c$ in \cite{Han93}, we use the energy shift between $^9$Li$_\mathrm{g.s.}$ and
$^9$Be$_\mathrm{IAS}$, which, as shown in Fig.~\ref{levels}, is equal to 0.7855 MeV. All these values, together with the
overlap of the spatial IAS and AAS wave functions, computed to be 0.74, and the core and halo
isospins, $t_c$ and $t_h$, leads to the expectation value $\langle H_\mathrm{coupl} \rangle =0.35$ MeV,
which by means of Eq.(8) in \cite{Han93}, and making use of the computed energy difference between the IAS and
AAS, 5.77 MeV, leads to an isospin mixing $\alpha^2=\sin^2\theta=3.7 \times 10^{-3}$.

\subsection{Gamow-Teller decay.}

The Gamow-Teller decay matrix element is given by Eq.(\ref{eq4}), and it contains the summation of two terms: the 
matrix element of the Gamow-Teller operator involving the valence nucleons, and a term proportional to the same overlap,
$\langle \Psi_{3b}^\mathrm{(fin)}  || \Psi_{3b}^\mathrm{(in)} \rangle$, 
as the one needed for the Fermi decay, and which
comes from the Gamow-Teller decay of the core. The mixing of these two terms is dictated by the factor
\begin{equation}
f_{\mathrm{mix}}= f_{\mathrm{fin}}
\langle t_c, t_c^z | \sum_{j \in  \mbox{ \scriptsize $^9$Li}}  \bm{\sigma}_j t_j^+ |  t'_c=\frac{3}{2}, 
t^{\prime z}_c=-\frac{3}{2} \rangle,
\label{fmix}
\end{equation}
which includes the Gamow-Teller matrix element describing the decay of the $^9$Li$_{\mathrm{g.s.}}$ core
into a core with isospin quantum numbers $t_c$ and $t_c^z$.

The first of the two terms mentioned above, the matrix element describing the Gamow-Teller decay of the valence nucleons,
is zero for decays into $^{11}$Be$_{\mathrm{g.s.}}$. This is due to the different core isospin in the initial
($t'_c=\frac{3}{2}$) and final ($t_c=\frac{1}{2}$) states. For decays into $^{11}$Be$_{\mathrm{IAS}}$ and $^{11}$Be$_{\mathrm{AAS}}$ this is not true anymore, because, as seen
in Eqs.(\ref{isoIAS}) and (\ref{isoAAS}), the wave functions for these two states have a component corresponding to a $^9$Li$_{\mathrm{g.s.}}$ core plus a neutron
and a proton. However, the numerical calculation of this matrix element for these two transitions gives rise to 
a very small number, about $10^{-3}$ in both cases, which permits to predict that the contribution to the Gamow-Teller 
strength from the decay of one of the halo neutrons is going to be, in any case, very small. This is due to the
fact that this contribution is given only by the components in the initial and final wave functions with relative nucleon-nucleon $p$-waves and nucleon-nucleon spin equal to 1, which have a very minor weight in the wave functions.

\begin{figure}
\centering
\includegraphics[width=\linewidth]{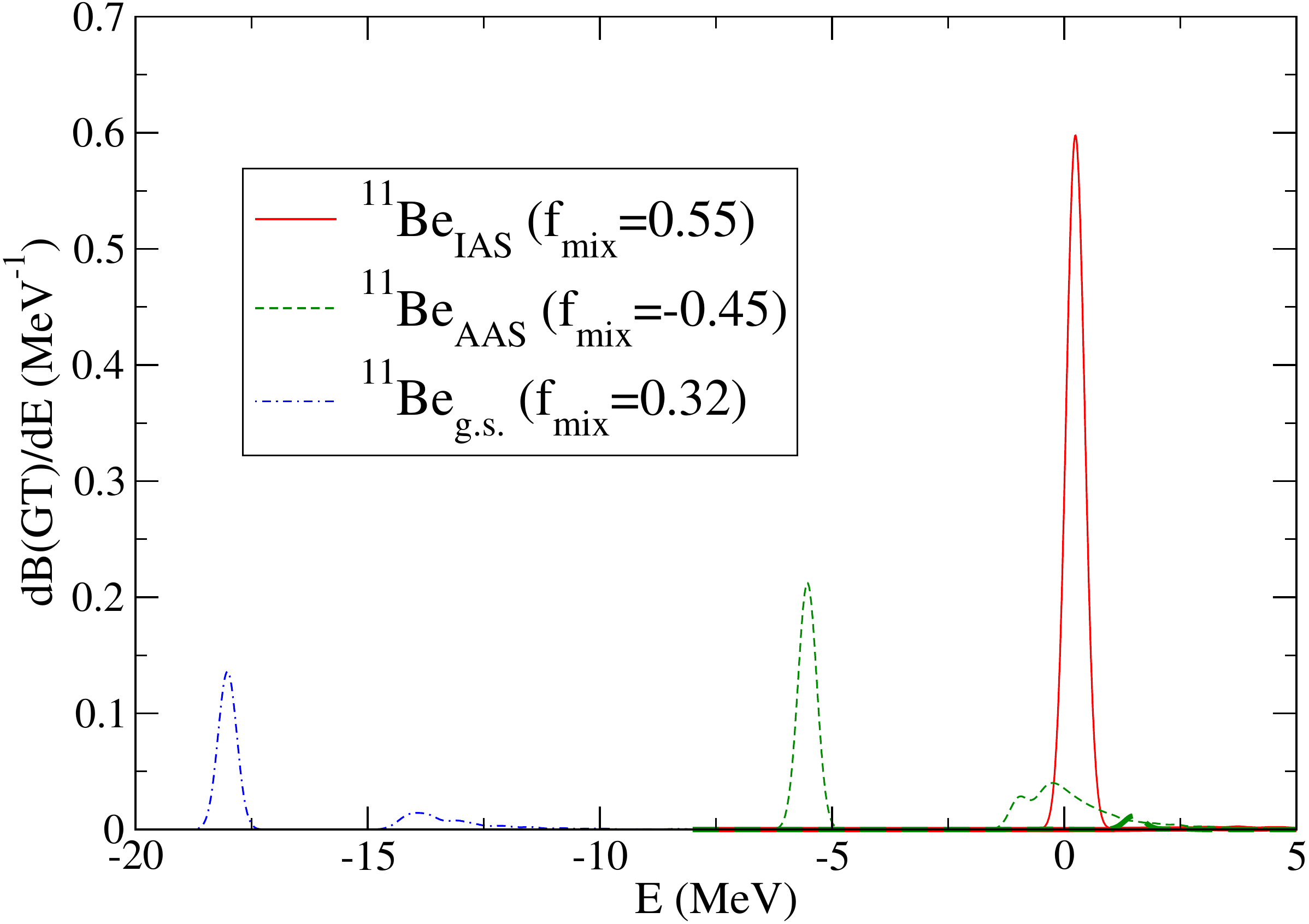}
\caption{Gamow-Teller strength, Eq.(\ref{dstr}), for decay of $^{11}$Li$_{\mathrm{g.s.}}$ into states,
bound and continuum, with the same quantum numbers as $^{11}$Be$_{\mathrm{IAS}}$ (solid curves), 
$^{11}$Be$_{\mathrm{AAS}}$ (dashed curves), and
$^{11}$Be$_{\mathrm{g.s.}}$ (dot-dashed curves), as a function of the energy of the final state. 
The mixing factor, $f_\mathrm{mix}$, is equal to 0.55, $-0.45$, and 0.32, for each of the
three cases, respectively. The hardly visible thick curves in the low-right corner show
the very small contribution from the decay of the halo neutrons.}
\label{fig2}
\end{figure}

The second term in the Gamow-Teller strength comes from beta decay of the core, whose beta decay
transition matrix element enters in the parameter given in Eq.(\ref{fmix}). From Eqs.(\ref{ffin2}) and
(\ref{ffin3}) we see that for both, the decay into $^{11}$Be$_{\mathrm{IAS}}$ and $^{11}$Be$_{\mathrm{AAS}}$,
the core decay matrix element describes the decay $^9$Li$_\mathrm{g.s.}$ into the same core 
state, in particular the core state with $t_c=\frac{3}{2}$ and $t_c^z=-\frac{1}{2}$, which 
corresponds to $^9$Be$_\mathrm{IAS}$. In the same way, for a transition
into $^{11}$Be$_\mathrm{g.s}$, we see from Eq.(\ref{ffin1}) that $t_c=\frac{1}{2}$ and $t_c^z=-\frac{1}{2}$,
and the core matrix element in Eq.(\ref{fmix}) refers to the decay of $^9$Li$_\mathrm{g.s.}$ into
$^9$Be$_\mathrm{g.s.}$.

Appropriate values for the core Gamow-Teller matrix elements may be estimated from shell-model calculations 
for the decay of $^{9}$Li$_\mathrm{g.s.}$, as shown in Ref.~\cite{Suz97}. In this work the values of 0.5 and 0.1 are given for
the Gamow-Teller strength for decays of $^9$Li$_\mathrm{g.s.}$ into $^9$Be$_\mathrm{IAS}$ and $^9$Be$_\mathrm{g.s.}$,
respectively. The square root of these values gives then the transition matrix element entering in Eq.(\ref{fmix})
for each case, which give rise to $f_\mathrm{mix}=0.55$, $-0.45$, and 0.32
for decays into $^{11}$Be$_\mathrm{IAS}$, $^{11}$Be$_\mathrm{AAS}$, and $^{11}$Be$_\mathrm{g.s.}$, respectively.
To get these numbers one has to recall that $f_\mathrm{fin}$ is equal to $\sqrt{3/5}$, $-\sqrt{2/5}$, and 1
for each of the transitions mentioned above (third column in Table~\ref{tobis}).

\begin{table}
\begin{tabular}{l|ccc}
    &   $^{11}$Be$_\mathrm{IAS}$ & $^{11}$Be$_\mathrm{AAS}$ &  $^{11}$Be$_\mathrm{g.s.}$ \\ \hline
Fermi  & 5.0  &  0   & 0    \\ \hline
Gamow-Teller   & 0.30  &  0.18   & 0.09   \\ 
GT (no cont.)  &  $0.29$ & 0.10 & 0.07    \\
GT (continuum) &  $<0.01$ & 0.08 & 0.02  \\
$f_\mathrm{mix}$  &  0.55 &  $-0.45$  &  0.32  \\ \hline
\end{tabular}
\caption{Integrated strengths for Fermi and Gamow-Teller decay of $^{11}$Li$_\mathrm{g.s.}$ into 
$^{11}$Be$_\mathrm{IAS}$, $^{11}$Be$_\mathrm{AAS}$, and $^{11}$Be$_\mathrm{g.s.}$. For the case
of Gamow-Teller decay we split the total strength into the contributions from the non-continuum states and the continuum background. The last row gives the$f_\mathrm{mix}$ values, Eq.(\ref{fmix}), used to compute the Gamow-Teller
strengths. }
\label{tab3}
\end{table}

Using the $f_\mathrm{mix}$ values given above, we obtain the computed Gamow-Teller strengths, Eq.(\ref{dstr}), shown in 
Fig.~\ref{fig2} for decay of $^{11}$Li$_{\mathrm{g.s.}}$ into states,
bound and continuum, with the same quantum numbers as $^{11}$Be$_{\mathrm{IAS}}$ (solid curves), 
$^{11}$Be$_{\mathrm{AAS}}$ (dashed curves), and $^{11}$Be$_{\mathrm{g.s.}}$ (dot-dashed curves).
The thick curves in the low-right corner of the figure show the 
tiny contribution to the strength from the beta decay of the halo neutrons (which is exactly zero for the case of
decay into $^{11}$Be$_\mathrm{g.s.}$). This contribution is hardly visible, and the consequence is that 
the curves in Fig.~\ref{fig2} are simply the ones in Fig.~\ref{fig1} scaled
by the factor $f_\mathrm{mix}^2$. As a consequence, the integrals under the curves is nothing but 
the product of $f_\mathrm{min}^2$ and the values of $B_\mathrm{ov}$ given in Table~\ref{tab2}. In this way
we get the total Gamow-Teller strengths 0.30, 0.18, and 0.09 for transitions into $^{11}$Be$_\mathrm{IAS}$, $^{11}$Be$_\mathrm{AAS}$, and $^{11}$Be$_\mathrm{g.s.}$, respectively, from which 0.29, 0.10, and 0.07
correspond to transitions into the non-continuum IAS, AAS, and $^{11}$Be$_\mathrm{g.s.}$, and the rest
is given by decay into the continuum. These results have been collected into the lower part of Table~\ref{tab3},
where the last row gives the $f_\mathrm{mix}$ values used, and where, for completeness, 
we also give the strengths for Fermi decay.

\section{Comparison to data}
\label{sec:comparison}

A brief overview of the current experimental state of knowledge on the beta decay of $^{11}$Li and its core $^9$Li is given in Appendix~\ref{app4} and Fig.~\ref{fig:li9_11}. Here we emphasize those aspects that are relevant to compare with our calculations.

\subsection{Three-body energies}

Our calculations should be most reliable for the $3/2^-$ IAS and AAS. The IAS energy has been adjusted to the experimental position at 21.16 MeV as determined in a charge-reaction experiment \cite{Ter97}, but neither the total strength nor the isospin purity could be accurately extracted in the experiment. The AAS position is then predicted to be at 15.39 MeV (see Table~\ref{taben}), which fits naturally with the experimentally established $3/2^-$ level at 16.3 MeV.

\subsection{The IAS and AAS strengths}

The Fermi strength of 5 to the IAS is significantly larger than the predicted Gamow-Teller strength of 0.29 and will dominate the transition. The transition is not energetically allowed in beta decay and it has not been possible to extract an accurate strength from the existing charge-reaction experiment.
The predicted GT beta strength to the AAS is 0.10 and thereby close to the so far identified experimental strength of 0.12. We note that not all experimental strength may have been identified yet and that the Fermi strength is negligible.
The predicted isospin impurity of the IAS, $3.7 \times 10^{-3}$, is
large for such a light nucleus, but still too small to be easily
measured. The magnitude is, as is the mismatch of 0.02 in the 
spatial overlap $^{11}$Li to IAS of 0.98, increased due to the halo structure.

\begin{figure}
\centering
\includegraphics[width=\linewidth]{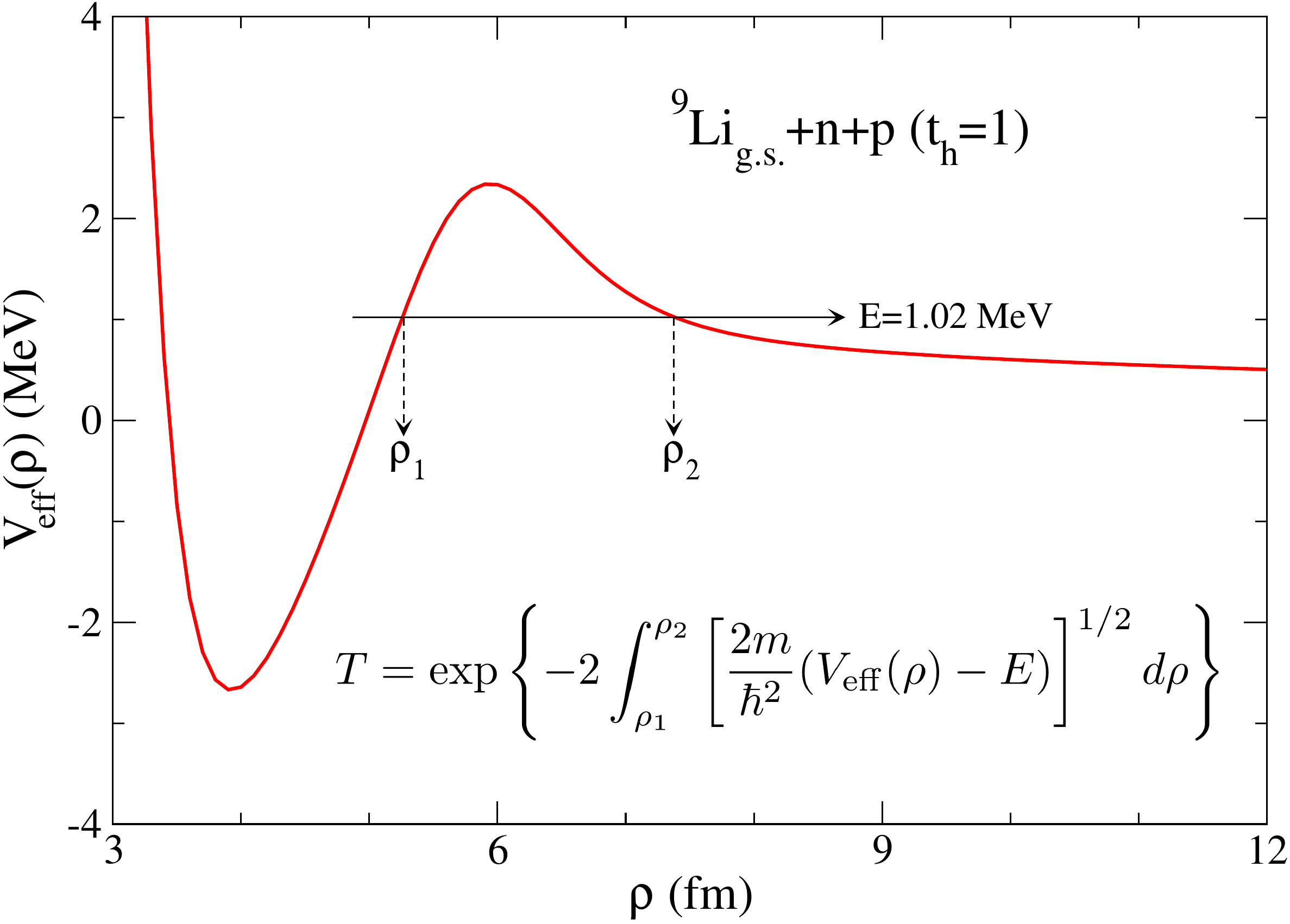}
\caption{Lowest effective potential for the $^9$Li$_\mathrm{g.s.}$+$n$+$p$ system with $t_h=1$, and turning points, $\rho_1$ 
and $\rho_2$, for the energy of the $^{11}$Be$_\mathrm{IAS}$ state. The WKB transmission coefficient through the barrier, $T$,
is obtained as indicated in the figure.}
\label{figwkb}
\end{figure}

\subsection{Decay branches}

As shown in Eq.(\ref{isoIAS}), the IAS contains two different three-body components, the $^9$Be$_\mathrm{IAS}$+$n$+$n$
and the $^9$Li$_\mathrm{g.s.}$+$n$+$p$ structures, both with $t_h=1$. As seen in Fig.~\ref{levels}, the IAS energy
is below the $^9$Be$_\mathrm{IAS}$+$n$+$n$ threshold by 0.55 MeV, but above the $^9$Li$_\mathrm{g.s.}$+$n$+$p$ threshold
by 1.02 MeV. Therefore, whereas the IAS can not decay through the $^9$Be$_\mathrm{IAS}$+$n$+$n$ channel,
it actually can do it through the $^9$Li$_\mathrm{g.s.}$+$n$+$p$ channel. An estimate of the width for this decay
can be obtained by means of the WKB approximation, as described in Ref.~\cite{gar04b}. In Fig.~\ref{figwkb} we show the lowest
effective potential for the $^9$Li$_\mathrm{g.s.}$+$n$+$p$ ($t_h=1$) system, obtained with potential $i)$, given in 
Table~\ref{tab1}, and indicate the turning points, $\rho_1$ and $\rho_2$, corresponding to the IAS excitation energy.
The transmission coefficient through the barrier, $T$, is obtained as indicated in the figure. The width, $\Gamma$, is then
estimated as $\Gamma/\hbar=fT$, where the knocking rate, $f$, is obtained as $f=v/\rho_1$, and the velocity $v$ is extracted
after equating the kinetic energy, $m v^2/2$, and the energy difference between $E$ and the maximum depth of the effective
potential. Following this procedure we have obtained a WKB estimate for the width of the IAS of about 0.5 MeV, which
agrees nicely with the experimental width, Ref.~\cite{Ter97}, of 0.49(7) MeV.

\begin{figure*}
\centering
\includegraphics[width=16cm]{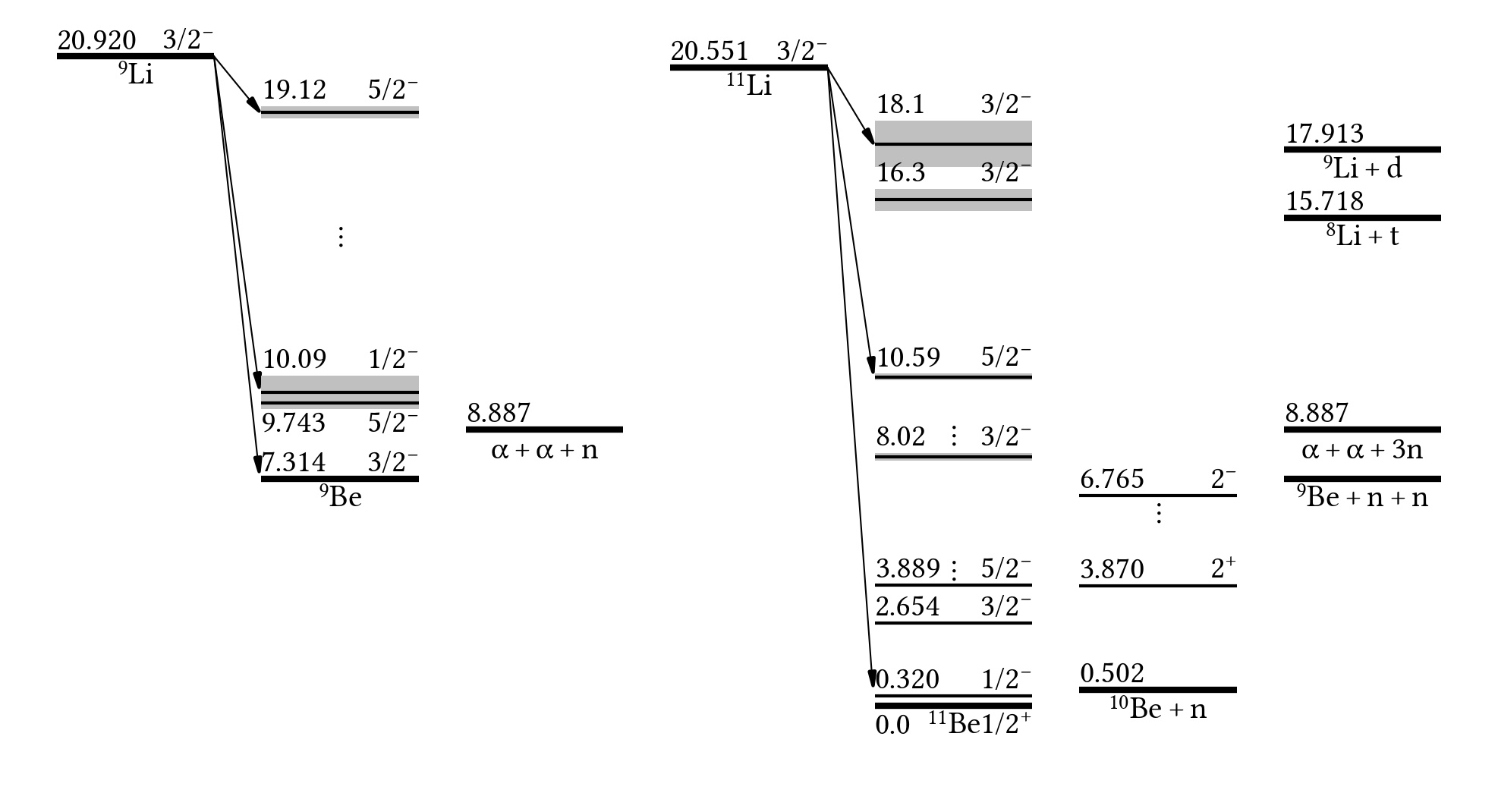}
\caption{Comparison of the beta decays of $^9$Li and $^{11}$Li. A selection of states fed in the decays is shown along with important thresholds for particle emission.}
\label{fig:li9_11}
\end{figure*}

We do not predict strong particle decay branches from our AAS state, but as remarked earlier it is likely to mix with close-lying states of the same spin-parity. We therefore expect its decay pattern to be very similar to that of the 18 MeV level, an expectation that agrees with the current experimental knowledge \cite{Mad09a}.

Moving now to the $^{11}$Be$_\mathrm{g.s.}$ calculation, they are
based on a shell-model strength for the $^{9}$Be ground state
transition that is a factor 4 above the experimental value. Our above
predictions of a Gamow-Teller strength of 0.07 to a level at 2.89 MeV
excitation energy in $^{11}$Be and a strength of 0.02 situated around
the 2$n$-threshold (see Table~\ref{tab3}) should therefore be scaled down and are then below
the experimental strength of 0.075 to the $3/2^-$ levels at 2.69 MeV and 
3.97 MeV. We shall return below to the question of what the origin of
the strength to the low-lying states in $^{11}$Be is.
Still, we note that Fig.~\ref{fig2} predicts strength to the states
just below the one neutron threshold in $^{10}$Be as well as strength
going directly to the two-neutron continuum. The barely bound states
are experimentally known to
be fed and have been proposed to have halo structure causing them to
be preferentially fed in the $^{11}$Li decay, see \cite{Mat09} and
references therein. Here again the experimental strength points to
the presence of other contributions.

\subsection{Core decay}

Model calculations for core decays into excited $^{9}$Be states would require five-body calculations and are not currently doable. The following observations are therefore of a more speculative nature:
Both decaying nuclei, $^9$Li and $^{11}$Li have large Q-values with low particle separation energies in the daughter nuclei and final states that in many cases can be considered composed of alpha particles and neutrons. The observable beta-strength will be
concentrated at high excitation energy and therefore clearly in the continuum. The spatial extension of the halo neutrons will tend to have larger overlap with final nucleons at low energy whereas the excitation energy residing in the core will be higher.
The (final state) interactions between nucleons and alpha particles are strong at relative energies up to a few MeV, so it is not surprising that final states in the $^9$Li decay tend to cluster into intermediate $^8$Be+n and $^5$He+$\alpha$ states, see e.g.\ figure 1 in \cite{Pre05}. A similar clustering has been found \cite{Mad08} for $^{11}$Li decays into two charged fragments where final state $^4$He+$^7$He structures have been clearly seen and there is evidence also for a $^6$He(2$^+$)+$^5$He channel; both configurations appear by adding two neutrons to one fragment in the $\alpha$+$^5$He channel.
The different final channels may be difficult to distinguish at low total energies, so the first few MeV above the two-neutron threshold in $^{11}$Be could be challenging experimentally as well as numerically.

The first excited state in $^9$Be fed in beta decay is the narrow (0.78(13) keV) $5/2^-$ level at 2.43 MeV where the $B$(GT) is 0.046(5) \cite{til04}. This could correspond to the $5/2^-$ 10.6 MeV level in $^{11}$Be (see Fig. \ref{fig:li9_11}) to which the branching ratio seems to be around 2\% (adding the values in \cite{Mat09,Mad08}, the branching ratio from \cite{Hir05} is higher, but is inconsistent with the data of \cite{Mad08}) giving a $B$(GT) around 0.1. 

The second state fed in $^9$Be is the very broad (1.08(11) MeV)
$1/2^-$ level at 2.78 MeV. It is the only $1/2^-$ level fed in the
$^9$Li decay and has a strong $2\alpha+$n clustering. A $^{11}$Li beta
decay with this ``core component'' would naturally lead to
$2\alpha+3$n final configurations. The higher-lying states fed in
$^9$Be are also broad and would also be expected to lead to high-lying
multi-body final states. As a rough indicator, the partial half life
for all excited states in $^9$Be --- all leading to $2\alpha+$n--- is
around 350 ms, which would correspond to a $^{11}$Li branching ratio
of 2.5\% that is close to $P_{3n} = 1.9(2)$\%. This lends support to
the suggested parallel between the $^9$Li and $^{11}$Li decays.

The main fact speaking against this suggested parallel is that the spin-parity of the strongly fed 11.8 MeV state in $^9$Be is $5/2^-$, whereas the corresponding strongly fed 18 MeV state in $^{11}$Be is reported as $3/2^-$, see figure \ref{fig:li9_11}. Both spin determinations come from angular distributions assuming the decay channels could be clearly separated. If this discrepancy is confirmed when a better decay scheme is established, a core decay scenario would be excluded.

\subsection{Decay into $t_h=0$ components}

Finally, let us consider the additional halo decay components. The IAS and AAS correspond to $t_h=1$ final state components.
Decay into the $t_h=0$ components ($T=\frac{3}{2}$, $T_z=-\frac{3}{2}$) will include core-deuteron structures. These states
can be populated after Gamow-Teller beta decay of $^{11}$Li$_\mathrm{g.s.}$, and the corresponding strength
will be given by the first term in Eq.(\ref{eq4}) only, i.e., the daughter nucleus is populated by decay of one of
the halo neutrons in $^{11}$Li$_\mathrm{g.s.}$.

\begin{figure}
\centering
\includegraphics[width=\linewidth]{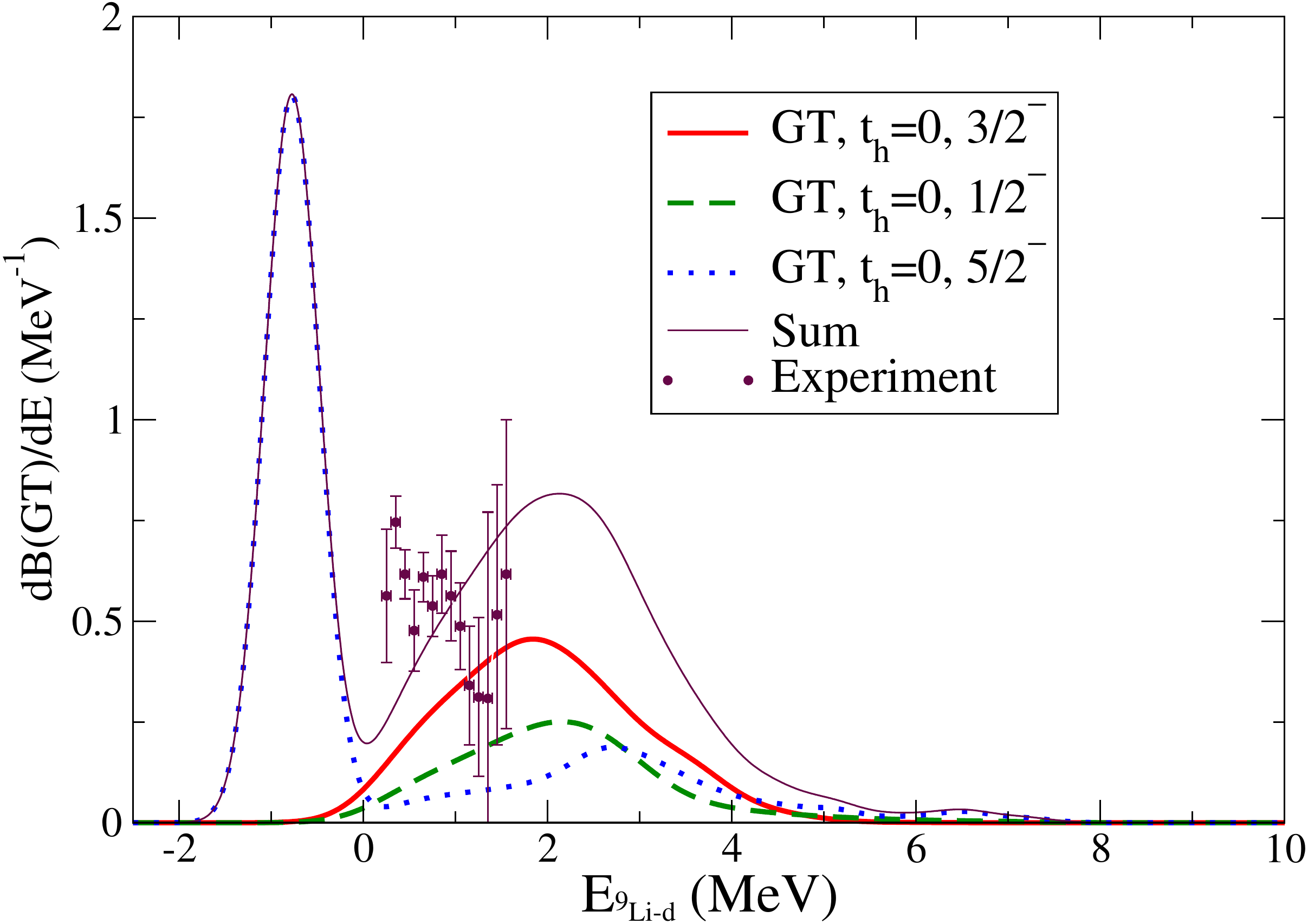}
\caption{Gamow-Teller strength, Eq.(\ref{dstr}), for decay of $^{11}$Li$_{\mathrm{g.s.}}$ into $t_h=0$ states in 
$^{11}$Be with total angular momentum and parity $J^\pi=\frac{3}{2}^-$ (solid), $J^\pi=\frac{1}{2}^-$ (dashed),
and $J^\pi=\frac{5}{2}^-$ (dotted). The sum of the three terms is shown by the brown (thin-solid) curve. The strength 
is given as a function of the relative $^9$Li$_\mathrm{g.s.}$-deuteron energy. The experimental data are from \cite{Rii22,Raa08}.}
\label{th0}
\end{figure}

In Fig.~\ref{th0} we plot the Gamow-Teller strength for decay of $^{11}$Li$_{\mathrm{g.s.}}$ into $t_h=0$ states in 
$^{11}$Be with total angular momentum and parity $J^\pi=3/2^-$ (solid), $J^\pi=1/2^-$ (dashed),
and $J^\pi=5/2^-$ (dotted).  In the case of $J^\pi=5/2^-$ there is a $^{11}$Be state about 0.8 MeV below the 
$^9$Li$_\mathrm{g.s.}$+deuteron threshold, which is responsible for the large peak observed in the dotted
curve.
For the other two cases, $J^\pi=1/2^-$ and $J^\pi=3/2^-$, all the computed 
states are found above the deuteron threshold. The sum of the three terms is shown by the thin-solid curve.
 The total integrated Gamow-Teller strengths are 1.21, 0.65, 
and 1.88 for the $3/2^-$, $1/2^-$, and $5/2^-$ cases, respectively.
There is at the moment no experimental evidence for the $5/2^-$ level
below the threshold, but decays directly to the deuteron continuum have
been seen, the latest experiment \cite{Rii22,Raa08} observing deuterons in
the energy range 0.2 MeV to 1.6 MeV above the deuteron threshold with
a total experimental $B$(GT) of 0.7 \cite{Rii22}. These experimental data
are shown by the dots in the figure. The predicted
integrated strengths in this range are 0.52, 0.25, and 0.11
for the $3/2^-$, $1/2^-$, and $5/2^-$ cases, respectively. The sum of
0.88 is slightly above the experimental value and the theoretical
strength increases with energy whereas the experimental is close to constant.

\subsection{Decays outside the model}

What remains is the decays that are outside our model space, namely
the final states containing proton $p_{3/2}$ states fed by decay of a
$p_{1/2}$ neutron. The $\bm{\sigma}$ operator favours the $p_{1/2}$ to
$p_{3/2}$ transition to $p_{1/2}$ to $p_{1/2}$. For a simple estimate,
the existing proton in $^{11}$Li will block one out of the four
$p_{3/2}$ substates and the two protons will couple to $0^+$ or $2^+$,
these states in $^{10}$Be can then couple with the remaining $p_{1/2}$
neutron to give $1/2^-$ or $3/2^-$ and $5/2^-$ states,
respectively. The former could explain the feeding to the first
excited $^{11}$Be state \cite{Suz94}, the latter will contribute to the excited states up to around 9 MeV.

We have calculated the one-particle wave function overlap between a 
$^{11}$Li $p_{1/2}$ neutron and a well-bound proton or the halo neutron in the 
first excited $^{11}$Be state to be 0.75 and 0.96, respectively. The $(p_{1/2})^2$
fraction in $^{11}$Li is 0.35, the $p_{1/2}$ to $p_{3/2}$ fraction is 8/9, 
adding the Pauli blocking, the Gamow-Teller quenching and the wave function 
overlap all in all reduces the total halo strength of 6 to around 0.9. 
The sum of the currently assigned experimental strength to this region is around 0.6, which 
is in fair agreement.

\section{Conclusion}

We have extended the theoretical description of three-body systems to
include isospin explicitly, thereby allowing a treatment of beta-decay
processes. The framework reproduces selected parts of the decay well,
in particular the decay to the isobaric analogue state. It is also
expected to reproduce the main features of the decay to the
anti-analogue state that, however, can mix with surrounding levels. A
distinct advantage is the ability to treat decays (both weak and
strong) to final continuum states thereby giving a full description of
the decay process. This is important for decays of halo nuclei. We
have applied the framework to the decay of $^{11}$Li that is a
challenging case due to the non-zero spin of the core, $^9$Li, and to
the many relevant final-state channels.

Our description of both the IAS and AAS, where we identify the latter as the 16.3
MeV resonance, in $^{11}$Be is consistent with all existing
experimental data. We adjust the position of the IAS and then predict
the position of the AAS to be close to the experimental one. The decay
width of the IAS matches the experiment, and the decay pattern of the
AAS is close to that of the nearby 18 MeV level, as could be
expected. We note that the different potentials for core nucleon
systems with $t_{cN} = 2$ and $t_{cN} = 1$ have been fitted
independently, but that the strengths of the central potentials for
different angular momenta are consistent with the difference being due
to a common isospin-isospin nucleon-nucleon potential, as discussed in Appendix~\ref{app3}. 
Our predicted isospin mixing for the IAS of $3.7 \times 10^{-3}$ is
very large for a light nucleus and significantly larger than expected
for more bound nuclei.

The core internal degrees of freedom are only taken crudely into
account through the matrix elements for decay of $^9$Li into $^9$Be
states, and we furthermore only calculate explicitly results for
decays into the $^9$Be ground state. Nevertheless, a consistent
overall interpretation of the beta decay of $^{11}$Li emerges as
follows: the core beta decay explains much of the structure from around the two-neutron threshold up to the Q-value. The decay to the ground state of $^9$Be has low intensity (actually, the sum of $B$(GT) values to the lowest 9 MeV in $^9$Be is of order 0.1, only) and will give a small contribution to the $^{11}$Li decays to the lowest 9 MeV in $^{11}$Be that has a summed $B$(GT) of order 0.6. These decays get a larger contribution from the decay of the $p_{1/2}^2$ halo component where a produced $p_{3/2}$ proton couples with the core to make intermediate $^{10}$Be $0^+$ and $2^+$ states.
Apart from the transition to the AAS mentioned above and the
transition to the 18 MeV state that is assumed to correspond to the
$^9$Li transition to the 11.8 MeV state (note however the need for a
reinvestigation of this particular branch), the other high beta
strength transition at high excitation is the beta-delayed deuteron
branch that is fairly well reproduced as a halo beta decay.

The large Q$_{\beta}$ value for $^{11}$Li is explained by the fact that adding two neutrons to $^9$Li produces a barely bound nucleus, whereas adding two neutrons to $^9$Be gives a significant energy gain, see Fig. \ref{fig:li9_11}. The gain comes mainly from the first neutron that fills the $p_{3/2}$ neutron orbit, whereas the second neutron again produces barely bound states. The above interpretation of where core and halo decays proceed to nicely fit this observation.

There are several clear consequences of this interpretation that can be tested experimentally. In general, the spectator halo neutrons from core decays should give low-energy final state neutrons, this is consistent with the results from the two-neutron coincidence spectrum in \cite{Del19} where the majority of the coincident neutrons lie below 2 MeV. A specific requirement is that the 11.8 MeV state in $^9$Li must have spin $3/2^-$ rather than the current experimental value of $5/2^-$.

There is a need for improved experimental data on both Li-decays. The current discrepancies for the decay scheme of $^{11}$Li must be resolved and more detailed data on decays feeding the two- and three-neutron unbound continuum would be very valuable, although challenging to obtain as neutron detection is difficult. For $^9$Li it is important to look closer at the strength residing above 10 MeV, both in order to resolve the puzzle of the spin of the 11.8 MeV level and in order to determine whether other levels contribute as predicted by several models. A renewed look at and comparison with the mirror decay of $^9$C may be useful.

The beta decay of $^{11}$Li is challenging for several reasons. It would be interesting to test the methods developed here to decays of other dripline nuclei that can be described with few-body methods.

\begin{acknowledgments}
  This work has been partially supported by the Independent Research Fund Denmark (9040-00076B),
  by grant No.PGC2018-093636-B-I00, funded by MCIN/AEI/10.13039/501100011033 and
by ``ERDF A way of making Europe''.
\end{acknowledgments}

\appendix

\section{Isospin and interactions}
\label{app3}

The basic nucleon-nucleon interaction contains isospin-isospin terms, e.g.\ from one-pion exchange, and as Lane \cite{lan62} pointed out this will for a system with one nucleon outside a core give an interaction term $(U/A_c)\bm{t} \cdot \bm{t}_c$ where the constant $U$ in mid-mass nuclei is of order 100 MeV. Robson \cite{rob65} showed how this naturally leads to isobaric analogue states of high purity, and also noted that the isospin mixing mainly arises from the region external to the nucleus.
These argument can be generalized to our case. The isospin-isospin interaction for a two-neutron halo will have the form $(U/A)(\bm{t}_2 \cdot \bm{t}_c+\bm{t}_3 \cdot \bm{t}_c + \bm{t}_2 \cdot \bm{t}_3)=(U/A)(\bm{t}_h \cdot \bm{t}_c+ \bm{t}_2 \cdot \bm{t}_3)$. Here the first term can be rewritten as $\bm{t}_h \cdot \bm{t}_c = 1/2(t_h^+t_c^-+t_h^-t_c^+)+t_h^zt_c^z$. It is straightforward to verify that the explicit charge-exchange terms will transform the two components of the wave functions
in Eqs.(\ref{isoIASgen}) and (\ref{isoAASgen}) into each other and thereby, in the absence of isospin-breaking terms in the Hamiltonian, enforce the structure of the IAS and the AAS.

The AAS is established in many light nuclei, e.g. in $^{12}$C, where it is situated at 12.71 MeV somewhat below the IAS at 15.11 MeV. It is known to play a role in both gamma and beta decays, see e.g. the contributions in \cite{wil69}, and often gives the main contribution to isospin purity violation \cite{zel17}.

The spatial and spin parts of the wave functions should have the same structure for the IAS and the AAS (the spatial extent can differ, as shown in Fig. ~\ref{figpots}b), and a substantial part of the difference in their energies will therefore come from the isospin-isospin interaction. For a two-neutron halo the
$\bm{t}_2 \cdot \bm{t}_3$ term gives the same contribution in the two states, so the relevant quantity is the expectation value of $\bm{t}_h \cdot \bm{t}_c$. This differs by $t_h+t_c$ between the AAS and the IAS. However, the effective value of $U$ will be less than for bulk nuclear matter: we are in light nuclei and furthermore are dealing with loosely bound structures, so since the nucleon-nucleon isospin-isospin interaction is of short range its average value will fall below a simple $1/A$ scaling. The value of $E_\mathrm{IAS}-E_\mathrm{AAS} = (U_\mathrm{eff}/A_c)(t_h+t_c)$ may be estimated from $^{12}$C. Scaling with mass number and isospin this would give a difference of about 7 MeV between the IAS and the AAS of $^{11}$Li, a value that should be taken as an upper limit since the effect of the halo extension is not considered. 

In our two-body potentials in Section~\ref{2bodypot} we have not employed any isospin dependent potential, but rather deduced potentials for each isospin sector. It is therefore interesting that the difference in the strengths for $t_{cN}=2$ and $t_{cN}=1$ in Table~\ref{tab1} is of similar order-of-magnitude for the central potential, consistent with the isospin dependence being effectively included.

\section{Potential matrix elements for 
$^{11}$Li$_{\mathrm{g.s.}}$, $^{11}$Be$_{\mathrm{g.s.}}$,
$^{11}$Be$_{\mathrm{IAS}}$, and $^{11}$Be$_{\mathrm{AAS}}$ }
\label{app1}

Here we specify how the matrix elements of the core-nucleon potentials in between the different terms of the basis 
set $\{|{\cal Y}_q\rangle |TT_z\rangle\}$ are computed for the particular cases of  $^{11}$Li$_{\mathrm{g.s.}}$, $^{11}$Be$_{\mathrm{g.s.}}$, $^{11}$Be$_{\mathrm{IAS}}$, and $^{11}$Be$_{\mathrm{AAS}}$.

From Eq.(\ref{tcnexp}) we immediately get that after rotation from the first to the second Jacobi set, the isospin wave functions
for the $^{11}$Li$_{\mathrm{g.s.}}$, $^{11}$Be$_{\mathrm{g.s.}}$,
$^{11}$Be$_{\mathrm{IAS}}$, and $^{11}$Be$_{\mathrm{AAS}}$ states is given by:
\begin{widetext}
\begin{equation}
|\mbox{$^{11}$Li}_{\mathrm{g.s.}} \rangle=|(t_2,t_3)t_h=1, t_c=\frac{3}{2} ; T=\frac{5}{2},T_z=-\frac{5}{2}\rangle=
                                                                        |(t_3,t_c)t_{cN}=2, t_2=\frac{1}{2} ; T=\frac{5}{2},T_z=-\frac{5}{2}\rangle,
                                                                        \label{rot1}
\end{equation}
\begin{equation}
|\mbox{$^{11}$Be}_{\mathrm{g.s.}} \rangle=|(t_2,t_3)t_h=1, t_c=\frac{1}{2} ; T=\frac{3}{2},T_z=-\frac{3}{2}\rangle=
                                                                        |(t_3,t_c)t_{cN}=1, t_2=\frac{1}{2} ; T=\frac{3}{2},T_z=-\frac{3}{2}\rangle,
                                                                        \label{rot1b}
\end{equation}
\begin{equation}
|\mbox{$^{11}$Be}_{\mathrm{IAS}} \rangle=|(t_2,t_3)t_h=1,t_c=\frac{3}{2} ; T=\frac{5}{2},T_z=-\frac{3}{2}\rangle=
                                                                         |(t_3,t_c)t_{cN}=2,t_2=\frac{1}{2} ; T=\frac{5}{2},T_z=-\frac{3}{2}\rangle,
                                                                         \label{rot2}
\end{equation}
\begin{eqnarray}
|\mbox{$^{11}$Be}_{\mathrm{AAS}} \rangle&=&|(t_2,t_3)t_h=1,t_c=\frac{3}{2}; T=\frac{3}{2},T_z=-\frac{3}{2}\rangle \nonumber \\
&=& \sqrt{\frac{5}{8}} |(t_3,t_c)t_{cN}=1,t_2=\frac{1}{2}; T=\frac{3}{2},T_z=-\frac{3}{2}\rangle -
         \sqrt{\frac{3}{8}} |(t_3,t_c)t_{cN}=2,t_2=\frac{1}{2}; T=\frac{3}{2},T_z=-\frac{3}{2}\rangle,
         \label{rot3}
\end{eqnarray}
\end{widetext}
where we can see that $^{11}$Li$_{\mathrm{g.s.}}$ and $^{11}$Be$_{\mathrm{IAS}}$ can only hold $t_{cN}=2$ values (otherwise the total isospin $T=\frac{5}{2}$ could not be reached) and the $^{11}$Be$_{\mathrm{g.s.}}$ can only have
 $t_{cN}=1$ (otherwise the total isospin $T=\frac{3}{2}$ could not be reached). On the contrary, $^{11}$Be$_{\mathrm{AAS}}$ mixes two different $t_{cN}$ values, $t_{cN}=1$ and $t_{cN}=2$.
 
 From the expressions above, and making use of Eq.(\ref{vcnmat}), we can split the matrix elements corresponding to the core-nucleon interaction into the different $t_{cN}$ parts:
\begin{widetext}
\begin{equation}
\langle {\cal Y}_q \mbox{$^{11}$Li}_{\mathrm{g.s.}} | V_{cN} |{\cal Y}_{q'} \mbox{$^{11}$Li}_{\mathrm{g.s.}} \rangle=
\langle{\cal Y}_q;  (t_3,t_c)t_{cN}=2,t_2=\frac{1}{2} ; \frac{5}{2},-\frac{5}{2}|  V_{cN}^{(t_{cN}=2)}  |{\cal Y}_{q'}; (t_3,t_c)t_{cN}=2,t_2=\frac{1}{2} ; \frac{5}{2},-\frac{5}{2}\rangle,
\label{matpot0}
\end{equation}
 \begin{equation}
\langle {\cal Y}_q \mbox{$^{11}$Be}_{\mathrm{g.s.}} | V_{cN} |{\cal Y}_{q'} \mbox{$^{11}$Be}_{\mathrm{g.s.}} \rangle=
\langle{\cal Y}_q;  (t_3,t_c)t_{cN}=1,t_2=\frac{1}{2} ; \frac{3}{2},-\frac{3}{2}|  V_{cN}^{(t_{cN}=1)}  |{\cal Y}_{q'}; (t_3,t_c)t_{cN}=1,t_2=\frac{1}{2} ; \frac{3}{2},-\frac{3}{2}\rangle,
\label{matpot0b}
\end{equation}
\begin{equation}
\langle {\cal Y}_q \mbox{$^{11}$Be}_{\mathrm{IAS}} | V_{cN} |{\cal Y}_{q'} \mbox{$^{11}$Be}_{\mathrm{IAS}} \rangle=
\langle{\cal Y}_q;  (t_3,t_c)t_{cN}=2,t_2=\frac{1}{2} ; \frac{5}{2},-\frac{3}{2}|  V_{cN}^{(t_{cN}=2)}  |{\cal Y}_{q'}; (t_3,t_c)t_{cN}=2,t_2=\frac{1}{2} ; \frac{5}{2},-\frac{3}{2}\rangle,
\label{matpot1}
\end{equation} 
and
\begin{eqnarray}
\langle {\cal Y}_q \mbox{$^{11}$Be}_{\mathrm{AAS}} | V_{cN} |{\cal Y}_{q'} \mbox{$^{11}$Be}_{\mathrm{AAS}} \rangle&=&
\frac{5}{8}
\langle {\cal Y}_q; (t_3,t_c)t_{cN}=1,t_2=\frac{1}{2} ; \frac{3}{2},-\frac{3}{2}|  V_{cN}  ^{(t_{cN}=1)}|{\cal Y}_{q'}; (t_3,t_c)t_{cN}=1,t_2=\frac{1}{2} ; \frac{3}{2},-\frac{3}{2}\rangle
\nonumber \\ &\!\!\!\!\!+&\!\!\!\!\!
\frac{3}{8}
\langle{\cal Y}_q;  (t_3,t_c)t_{cN}=2,t_2=\frac{1}{2} ; \frac{3}{2},-\frac{3}{2}|  V_{cN}^{(t_{cN}=2)}  |{\cal Y}_{q'};(t_3,t_c)t_{cN}=2,t_2=\frac{1}{2} ; \frac{3}{2},-\frac{3}{2}\rangle
\label{matpot2}
\end{eqnarray}
\end{widetext}
where we have assumed that the potential does not mix different values of the core-nucleon isospin, $t_{cN}$.

Finally, from Eq.(\ref{vt3tc}) we can find the specific core-nucleon states, and therefore the specific core-nucleon
interactions, involved in the different three-body systems
under consideration:
\begin{widetext}
\begin{equation}
\langle {\cal Y}_q; \mbox{$^{11}$Li}_{\mathrm{g.s.}} | V_{cN} |{\cal Y}_{q'}; \mbox{$^{11}$Li}_{\mathrm{g.s.}} \rangle=
\langle {\cal Y}_q; \mbox{$^9$Li$_{\mathrm{g.s.}}$}+n | V_{cN}^{(t_{cN}=2)} | {\cal Y}_{q'}; 
                                                           \mbox{$^9$Li$_{\mathrm{g.s.}}$}+n \rangle,
\label{9lin}
\end{equation}
\begin{equation}
\langle {\cal Y}_q; \mbox{$^{11}$Be}_{\mathrm{g.s.}} | V_{cN} |{\cal Y}_{q'}; \mbox{$^{11}$Be}_{\mathrm{g.s.}} \rangle=
\langle {\cal Y}_q; \mbox{$^9$Be$_{\mathrm{g.s.}}$}+n | V_{cN}^{(t_{cN}=1)} | {\cal Y}_{q'}; 
                                                           \mbox{$^9$Be$_{\mathrm{g.s.}}$}+n \rangle,
\label{9ben}
\end{equation}
\begin{eqnarray}
\lefteqn{ \hspace*{-1cm}
\langle {\cal Y}_q \mbox{$^{11}$Be}_{\mathrm{IAS}} | V_{cN} |{\cal Y}_{q'} \mbox{$^{11}$Be}_{\mathrm{IAS}} \rangle = \frac{1}{5}
\langle {\cal Y}_q; \mbox{$^9$Li$_{\mathrm{g.s.}}$}+n | V_{cN}^{(t_{cN}=2)} | {\cal Y}_{q'}; 
                                                           \mbox{$^9$Li$_{\mathrm{g.s.}}$}+n \rangle } 
                                              \nonumber             \\ & &
+\frac{1}{5} \langle {\cal Y}_q; \mbox{$^9$Li$_{\mathrm{g.s.}}$}+p | V_{cN}^{(t_{cN}=2)} | {\cal Y}_{q'}; 
                                                           \mbox{$^9$Li$_{\mathrm{g.s.}}$}+p \rangle
+\frac{3}{5} \langle {\cal Y}_q; \mbox{$^9$Be$_{\mathrm{IAS}}$}+n | V_{cN}^{(t_{cN}=2)} | {\cal Y}_{q'}; 
                                                           \mbox{$^9$Be$_{\mathrm{IAS}}$}+n \rangle,
\label{iasnp}
\end{eqnarray}
\begin{eqnarray}
\lefteqn{ \hspace*{-1cm}
\langle {\cal Y}_q \mbox{$^{11}$Be}_{\mathrm{AAS}} | V_{cN} |{\cal Y}_{q'} \mbox{$^{11}$Be}_{\mathrm{AAS}} \rangle = \frac{5}{32}
\langle {\cal Y}_q; \mbox{$^9$Be$_{\mathrm{IAS}}$}+n | V_{cN}^{(t_{cN}=1)} | {\cal Y}_{q'}; 
                                                           \mbox{$^9$Be$_{\mathrm{IAS}}$}+n \rangle } 
                                              \nonumber             \\ & &
+\frac{9}{160} \langle {\cal Y}_q; \mbox{$^9$Be$_{\mathrm{IAS}}$}+n | V_{cN}^{(t_{cN}=2)} | {\cal Y}_{q'}; 
                                                           \mbox{$^9$Be$_{\mathrm{IAS}}$}+n \rangle
+\frac{15}{32} \langle {\cal Y}_q; \mbox{$^9$Li$_{\mathrm{g.s.}}$}+p | V_{cN}^{(t_{cN}=1)} | {\cal Y}_{q'}; 
                                                           \mbox{$^9$Li$_{\mathrm{g.s.}}$}+p \rangle
                                              \nonumber \\  & &
+\frac{3}{160} \langle {\cal Y}_q; \mbox{$^9$Li$_{\mathrm{g.s.}}$}+p | V_{cN}^{(t_{cN}=2)} | {\cal Y}_{q'}; 
                                                           \mbox{$^9$Li$^{\mathrm{g.s.}}$}+p \rangle
+\frac{3}{10} \langle {\cal Y}_q; \mbox{$^9$Li$_{\mathrm{g.s.}}$}+n | V_{cN}^{(t_{cN}=2)} | {\cal Y}_{q'}; 
                                                           \mbox{$^9$Li$_{\mathrm{g.s.}}$}+n \rangle.
\label{aasnp}
\end{eqnarray}
\end{widetext}

Therefore, the required interactions are: $i)$ The interaction between
the $^9$Li$_{\mathrm{g.s.}}$ core and the neutron (only $t_{cN}=2$ is possible), $ii)$ the interaction
between the $^9$Be$_{\mathrm{g.s.}}$ core and the neutron (only $t_{cN}=1$ is possible),
$iii)$ the interaction between the $^9$Li$_{\mathrm{g.s.}}$ core and the proton for $t_{cN}=2$,
$iv)$ the interaction between the $^9$Be$_{\mathrm{IAS}}$ core and the neutron for $t_{cN}=2$,
$v)$ the interaction between the $^9$Li$_{\mathrm{g.s.}}$ core and the proton for $t_{cN}=1$,
and $vi)$ the interaction between the $^9$Be$_{\mathrm{IAS}}$ core and the neutron for $t_{cN}=1$.

\section{${\cal H}_{\mathrm{core}}$ matrix elements for 
$^{11}$Li$_{\mathrm{g.s.}}$, $^{11}$Be$_{\mathrm{g.s.}}$,
$^{11}$Be$_{\mathrm{IAS}}$, and $^{11}$Be$_{\mathrm{AAS}}$ }
\label{app2}

Following Eq.(\ref{hcgen}), we have that the effect of the core
hamiltonian, ${\cal H}_{\mathrm{core}}$, on the different core states involved in the three-body systems considered
in this work, can be expressed as:
\begin{equation}
{\cal H}_{\mathrm{core}} |t_c=\frac{3}{2},t_c^z=-\frac{3}{2}\rangle ={\cal H}_{\mathrm{core}} |\mbox{$^9$Li$_{\mathrm{g.s.}}$}\rangle=\xi_{\mbox{\scriptsize $^9$Li$_{\mathrm{g.s.}}$}} |\mbox{$^9$Li$_{\mathrm{g.s.}}$}\rangle,
\label{core1}
\end{equation}
\begin{equation}
{\cal H}_{\mathrm{core}} |t_c=\frac{1}{2},t_c^z=-\frac{1}{2}\rangle ={\cal H}_{\mathrm{core}} |\mbox{$^9$Be$_{\mathrm{g.s.}}$}\rangle=\xi_{\mbox{\scriptsize $^9$Be$_{\mathrm{g.s.}}$}} |\mbox{$^9$Be$_{\mathrm{g.s.}}$}\rangle,
\label{core1b}
\end{equation}
\begin{equation}
{\cal H}_{\mathrm{core}} |t_c=\frac{3}{2},t_c^z=-\frac{1}{2}\rangle = |\mbox{$^9$Be}_{\mathrm{IAS}}\rangle=\xi_{\mbox{\scriptsize $^9$Be}_{\mathrm{IAS}}} |\mbox{$^9$Be}_{\mathrm{IAS}}\rangle,
\label{core2}
\end{equation}
where $\xi_{\mbox{\scriptsize $^9$Li$_{\mathrm{g.s.}}$}}$,
$\xi_{\mbox{\scriptsize $^9$Be$_{\mathrm{g.s.}}$}}$,
 and $\xi_{\mbox{\scriptsize $^9$Be}_{\mathrm{IAS}}}$ are the energies of the $^9$Li$_{\mathrm{g.s.}}$,
$^9$Be$_{\mathrm{g.s.}}$,  and $^9$Be$_{\mathrm{IAS}}$ cores, respectively. 

Making now use of Eqs.(\ref{iso11Li}) to (\ref{isoAAS}), we can then easily get the matrix elements of ${\cal H}_{\mathrm{core}}$ between the different terms of the basis set:
\begin{equation}
\langle {\cal Y}_q \mbox{$^{11}$Li}_{\mathrm{g.s.}} | {\cal H}_{\mathrm{core}}  |{\cal Y}_{q'} \mbox{$^{11}$Li}_{\mathrm{g.s.}} \rangle=
 \xi_{\mbox{\scriptsize $^9$Li$_{\mathrm{g.s.}}$}} \delta_{qq'},
\label{matcor0}
\end{equation}
\begin{equation}
\langle {\cal Y}_q \mbox{$^{11}$Be}_{\mathrm{g.s.}} | {\cal H}_{\mathrm{core}}  |{\cal Y}_{q'} \mbox{$^{11}$Be}_{\mathrm{g.s.}} \rangle=
 \xi_{\mbox{\scriptsize $^9$Be$_{\mathrm{g.s.}}$}} \delta_{qq'},
\label{matcor0b}
\end{equation}
\begin{equation}
\langle {\cal Y}_q \mbox{$^{11}$Be}_{\mathrm{IAS}} | {\cal H}_{\mathrm{core}}  |{\cal Y}_{q'} \mbox{$^{11}$Be}_{\mathrm{IAS}} \rangle=
\left(
\frac{2}{5} \xi_{\mbox{\scriptsize $^9$Li$_{\mathrm{g.s.}}$}} +\frac{3}{5} \xi_{\mbox{\scriptsize $^9$Be}_{\mathrm{IAS}}}
\right) \delta_{qq'},
\label{matcor1}
\end{equation}
\begin{equation}
\langle {\cal Y}_q \mbox{$^{11}$Be}_{\mathrm{AAS}} | {\cal H}_{\mathrm{core}}  |{\cal Y}_{q'} \mbox{$^{11}$Be}_{\mathrm{AAS}} \rangle=
\left(
\frac{3}{5} \xi_{\mbox{\scriptsize $^9$Li$_{\mathrm{g.s.}}$}} +\frac{2}{5} \xi_{\mbox{\scriptsize $^9$Be}_{\mathrm{IAS}}}
\right) \delta_{qq'},
\label{matcor2}
\end{equation}
which are consistent with the general expressions given in Eqs.(\ref{coreh}), (\ref{coreIAS}), and (\ref{coreAAS}).
We have assumed that ${\cal H}_{\mathrm{core}}$ does not mix the $\mbox{$^{9}$Li}_{\mathrm{g.s.}}$ and 
$\mbox{$^{9}$Be}_{\mathrm{IAS}}$ wave functions.

\section{Decay data for $^9$Li and $^{11}$Li}
\label{app4}
Both decays involve broad (and partially overlapping) resonances in the daughter nuclei so that branching ratios and beta strength may not always be uniquely attributed to a given level. A discussion is given in \cite{Rii14} that suggests to consider $B$(GT) as a function of the final-state energy when comparing theory and experiment.
For narrow levels the usual relation between $B$(GT) and the $ft$-value is
$ft = 6144(4) / (B(\mbox{F}) + (g_A/g_V)^2B(\mbox{GT}))$ where the conversion factor is taken from superallowed Fermi transitions \cite{Har20} and neutron decay gives $g_A/g_V = -1.2754(13)$ \cite{pdg20}. The empirical quenching factor for Gamow-Teller strength was for light nuclei estimated in \cite{Cho93} to be 0.845 and 0.833 for mass 9 and 11 nuclei, giving effective values for $(g_A/g_V)^2$ of 1.16 and 1.13 that we shall use when extracting experimental $B$(GT) values.

\subsection{The $^9$Li decay}

A detailed overview of the current knowledge on the $^9$Li decay can be found from \cite{til04,Pre05}. It has historically been difficult to unravel the decays via excited $^9$Be states, see figure \ref{fig:li9_11}, since the obvious decay channels, through $\alpha$ + $^5$He and through n + $^8$Be, lead to the same final continuum and can overlap in energy. The currently established branches go to the $3/2^-$ ground state and excited states at 2.43 MeV ($5/2^-$), 2.8 MeV ($1/2^-$), 5.0 MeV ($3/2^-$), 7.9 MeV ($5/2^-$) and 11.8 MeV ($5/2^-$). The main Gamow Teller strength goes to the 11.8 MeV state, but has been difficult to determine accurately; different approaches give numbers in the range 5 to 8. The ground state transition has a strength of 0.026(1) and the states in between have strengths of similar order of magnitude. For the five lowest states this is all very consistent with earlier theoretical treatments \cite{Mik88, Cho93, Suz97, Kan10} that all predict several strong transitions in the energy range 10--13 MeV where only one has been observed.

\subsection{The $^{11}$Li decay}

A good overview of the existing data on the $^{11}$Li decay can be found in the review \cite{tunl11}, the only more recent experimental result being the first hint at the energy distribution for two-neutron coincidences \cite{Del19}. Most experiments have focused on the low-energy range, roughly up to around the two-neutron threshold in $^{11}$Be, and the overall features of the decay are known here. However, the detailed interpretation differs between the most recent experiments \cite{Mat09,Hir05} and there is still only few data from the higher-energy region, so the distribution of beta strength is still uncertain. Figure \ref{fig:li9_11} displays some of the levels that currently are believed to play a role in the decay.

Around 93\% of the decay goes to neutron unbound states, the bound state feeding goes to the first excited $1/2^-$ halo state in $^{11}$Be.
The measurement of Hirayama et al. \cite{Hir05} detected neutrons as well as gamma-rays and beta polarization. The extracted beta-asymmetry is for most neutron energies consistent with $^{11}$Be spins of $3/2^-$ or $5/2^-$ (or a mixture of both), it is only in very small regions at high neutron energy that $1/2^-$ is indicated with perhaps an indication of a contribution also a low neutron energies.
Their suggested level scheme differs in many details from the one extracted from the latest analysis \cite{Mat09} of $\beta$-$n$-$\gamma$ decays, but agree on including essentially the same $3/2^-$ and $5/2^-$ states up to 10.6 MeV. In terms of branching ratios they indicate that around 58\% of the decays go to states below the two-neutron threshold, and around 32\% to states above this up to the 10.6 MeV state. The total assigned probability for beta-delayed one-neutron emission does not exceed 70\%, which is considerably below the tabulated value of $P_{1n} = 86.3(9)$\%.

Only one $\beta$-$n$-$\gamma$ experiment \cite{Fyn04} reported neutrons from states at higher energies, the reported intensity of 6\% does not include neutron emission to the ground state nor the branching of around 1\% to n-$\alpha$-$^6$He. The multi-neutron emission probabilities of $P_{2n} = 4.1(4)$\% and $P_{3n} = 1.9(2)$\% combined with the small fractions of order $10^{-4}$ going via $\beta$d and $\beta$t shows that a sizeable strength (and a non-negligible branching) will go to the multi-particle continuum. The most recent measurements on branches with two charged particles \cite{Mad08,Mad09a} point to a significant role played by three levels at 10.6 MeV, 16.3 MeV and just above 18 MeV. The latter level has also been invoked in $\beta$t and $\beta$n branches, whereas the $\beta$d decay is normally assumed to proceed directly to the continuum. However, the overall experimental picture leaves many questions, in particular above the two-neutron threshold.

\end{document}